\pdfoutput=1

\documentclass[manuscript,nonacm]{acmart}
\setcopyright{none}

\AtBeginDocument{%
  }

\renewcommand\footnotetextcopyrightpermission[1]{}

\usepackage{graphicx}
\usepackage{subcaption}
\usepackage{booktabs}
\usepackage{multirow}
\usepackage{tabularx}
\usepackage{array}
\usepackage{longtable}
\usepackage{enumitem}
\usepackage{hyperref}
\usepackage{tikz}
\usetikzlibrary{arrows.meta,positioning}

\setlist[itemize]{leftmargin=2em, topsep=4pt, itemsep=2pt}
\setlist[enumerate]{leftmargin=2em, topsep=4pt, itemsep=2pt}
\newcolumntype{Y}{>{\raggedright\arraybackslash}X}

\begin{document}

\title{Afterglow: A Place-Based Memorial Ecology for AI-Mediated Pet Bereavement}

\author{Hanjing Shi}
\affiliation{%
  \department{Department of Computer Science and Engineering}
  \institution{Lehigh University}
  \city{Bethlehem}
  \state{Pennsylvania}
  \country{USA}}
\author{Dominic DiFranzo}
\affiliation{%
  \department{Department of Computer Science and Engineering}
  \institution{Lehigh University}
  \city{Bethlehem}
  \state{Pennsylvania}
  \country{USA}}
\renewcommand{\shortauthors}{Shi and DiFranzo}
\date{September 27, 2026}

\begin{abstract}
Pet bereavement often receives little social recognition. Generative AI can give a continuing bond a responsive voice, but a comforting reply may also claim authority to forgive or request attention. We investigate how a memorial can support connection without turning remembrance into obligation. Through Research through Design, we developed \textbf{Afterglow}, a mobile world connecting private remembrance, human witnessing, and symbolic Pet messages. A formative survey ($N=57$) informed the initial design, followed by six online roundtable walkthroughs with 20 unique participants across two prototype iterations. Our interpretive analysis develops tensions between \textit{returnable connection and emotional obligation}, \textit{recognizable likeness and ontological clarity}, and \textit{protective intervention and surveillant authority}. We contribute \textit{Legible Restraint}, a cross-layer requirement that limits on relational authority survive changes in speaker, generation context, trigger logic, data use, and participation. Its temporal consequence, \textit{Designing for Goodbye}, keeps remembrance available without making continued use a condition of care.
\end{abstract}



\maketitle
\hypersetup{pdfauthor={Hanjing Shi and Dominic DiFranzo}}

\section{Introduction}

A companion animal's death ends daily acts of care while the relationship may endure. Feeding, walking, and shared domestic routines may disappear, while attachment and the wish to remember remain~\cite{xiong2026bondi}. Yet caregivers can encounter dismissal, lack established mourning rituals, and feel pressure to justify why the relationship mattered~\cite{doka1989disenfranchised,cordaro2012pet,spain2019,park2023}. A digital memorial enters this gap between a bond that persists and everyday forms of care interrupted by death. Supporting that bond means giving mourners room to choose how remembrance fits into their lives.

HCI research has explored these needs through mixed-reality re-encounter, tangible memorials, reflective worlds, and online support groups~\cite{xiong2025petreunion_chi,xiong2026bondi,lefevre2024loss,kim2026oscillation}. Generative AI changes this encounter by turning photographs, stories, and remembered routines into replies written in the pet's voice. Research on human-loss chatbots shows that mourners can find such exchanges meaningful while recognizing their artificial source~\cite{xygkou2023conversation,campbell2025chatbots,stokes2026sensing}. Human responses also contribute voluntary attention and lived experience, forms of care distinct from an AI acknowledgement~\cite{meng2023mediated,li2026humanpeer}. A memorial system must accommodate these different relations and the expectations each brings.

The unresolved problem is how authority travels between those encounters. A symbolic Pet reply may become advice about guilt; a platform reminder may sound like the animal asking its caregiver to return. Here, \textit{Pet} names a generated system role, distinct from the deceased animal. We use \textit{relational authority} for the standing an interaction attributes to a speaker to comfort, judge, intervene, or request attention. A truthful AI label identifies the source without necessarily limiting that standing. First-person examples, automatic replies, and progression cues may assign responsibilities that the label appears to withhold. When absence seems to harm the Pet or diminish its memorial, an invitation to remember becomes a condition of care. We call this pressure \textit{coercive retention}.

Pet bereavement sharpens this coordination problem because the represented animal cannot authorize the voice inferred from caregiver-provided traces. Those traces may also belong to relationships among several caregivers. Moving a private photograph into generated speech or a shared postcard changes its use and audience~\cite{brubaker2014stewardship,lei2025aiafterlife,tran2025privacy}. How can private remembrance, human witnessing, symbolic exchange, and platform protection coexist while preserving their distinct sources of authority?

We investigate this question through \textbf{Afterglow}, a mobile memorial world developed through Research through Design (RtD)~\cite{zimmerman2007research}. The Cottage supports private remembrance, the River human exchange, the Forest small gestures of witnessing, and the Post Office sparse symbolic Pet messages. These independently accessible places offer several ways to return. A formative survey ($N=57$) informed the initial prototype. Six guided roundtable walkthroughs with 20 unique bereaved participants then examined two successive builds in a primarily Chinese-language research setting. This iterative inquiry concerns interpretations of memorial encounters and their design consequences.

The inquiry follows two questions from participants' needs to the artifact's response.

\begin{enumerate}
  \item What conflicting needs emerged around connection, likeness, and protection in an AI-mediated pet memorial?
  \item How did Afterglow's RtD iterations distribute relational authority across places and roles, and where did that separation become unstable?
\end{enumerate}

We argue that an AI memorial must be designed around what it may ask of the living. Our contribution, \textit{Legible Restraint}, treats relational authority as a cross-layer design requirement. Limits expressed through speaker identity and disclosure must also hold in generation context, trigger logic, data derivation, and exit paths. Three empirical tensions ground this requirement in conflicting demands for connection, likeness, and protection. Afterglow instantiates it through a place-based ecology, and three design tests trace where declared limits hold or break across an encounter. \textit{Designing for Goodbye} carries the same requirement through dormancy, return, and departure. This account shifts the design question from sustaining interaction to supporting a bond through changing patterns of participation.

\section{Related Work}

\subsection{Pet Loss, Disenfranchisement, and Continuing Bonds}

Pet loss is a paradigmatic form of disenfranchised grief, a meaningful loss that may be socially minimized or excluded from established mourning practices~\cite{doka1989disenfranchised,cordaro2012pet}. Reviews of pet bereavement describe loneliness, embarrassment, and limited social support. Research on pet-related guilt also shows how unrecognized caregiving expectations can themselves become disenfranchised~\cite{spain2019,park2023,kogan2022disenfranchised}. These experiences make recognition itself a design concern. A system matters by storing memories and by communicating that the relationship and its loss are legitimate.

Continuing-bonds theory replaces a simple opposition between healthy ``letting go'' and unhealthy attachment with relationships that persist and transform through memories, rituals, objects, and stories~\cite{klass1996continuing}. The dual process model describes bereavement as movement between loss-oriented and restoration-oriented coping, with respite forming part of that oscillation~\cite{stroebe1999dual}. Kim and Park apply this model to online pet-loss groups. Their interviews identify \textit{compulsory grief}, in which platform participation can hold people in a loss-oriented state, and motivate oscillation design between loss and restoration~\cite{kim2026oscillation}. A scoping review of 107 HCI papers organizes thanatotechnology around digital remains, remembrance, and coping, identifying practices that include curation, honoring, and letting go~\cite{albers2023dying}. Thanatosensitive design likewise calls for systems that respect the values and temporalities of death and approach grief on its own terms~\cite{massimi2009thanatosensitivity}. Work on domestic memorialization, interactive columbaria, technology heirlooms, and post-mortem stewardship develops material forms for repeated visiting, curation, and responsibility for digital remains~\cite{uriu2016fenestra,uriu2018columbaria,odom2012technology,brubaker2014stewardship,kim2024beside}. \textit{Living Memory Home} further locates continuing bonds in private ``backstage'' grieving, while film-based design fiction explores near-future memorial rituals through virtual funerals~\cite{she2021living,uriu2025virtualfunerals}. Sas et al. show that significant digital possessions can be both comforting and painful and argue for ritualized ways of letting them go~\cite{sas2016design}. Most artifacts in this lineage concern human death. Together, these practices show that continued remembrance, restorative activity, and deliberate disengagement can coexist as user-directed trajectories.

\subsection{Social and Culturally Situated Memorial Interaction}

When memorial activity moves online, continuing bonds become organized through interaction design. Online mourning expands the temporal, spatial, and social reach of remembrance, which becomes partly co-produced through profiles, comments, and audiences~\cite{brubaker2013beyond}. Massimi's ten-week deployment of an online bereavement group examined remembrance features and peer support within the same environment~\cite{massimi2013remembrance}. Research on sensitive disclosure further shows that support depends on audience response, reciprocity, and anonymity as well as on the act of disclosure~\cite{andalibi2018socialsupport}. Although this work primarily concerns human loss, it identifies mechanisms of visibility, reciprocity, and audience response that also shape socially minimized grief. In pet bereavement, being seen may validate both the grief and the relationship behind it, while silence or unwanted exposure can repeat the social minimization that made disclosure difficult. We therefore use this literature to frame the River as a remembrance and disclosure space whose audience and response conditions remain consequential.

These audience conditions are culturally situated. Research on digital mourning in China makes the relation among culture, platform structure, and interpersonal expectations especially visible. Mourning on WeChat Moments requires people to negotiate semi-public visibility, emotional restraint, and culturally appropriate expression~\cite{zhao2025wechatmourning}. On Weibo, mourners converse with deceased strangers, other mourners, and themselves, producing continuing bonds alongside social support~\cite{hu2025mourning}. Digital narratives can also reorganize disrupted rites of passage, while AI-mediated posthumous interaction introduces proxy governance, relational exposure, and negotiation within kinship~\cite{whyke2021rite,zhang2026voice}. These studies locate culture within a memorial system's interactional organization. Their relevance to Afterglow lies in how platform visibility and cultural norms organize witnessing, public grief, ritual, and authority. Pet bereavement brings these mechanisms into a relationship whose loss may itself require social recognition.

\subsection{Interaction Modalities and Temporal Commitments in Pet Memorial Systems}

Pet-specific HCI has approached continuing bonds through several primary forms of encounter. \textit{ReMember} is an interactive installation that combines heartbeat recordings, audiovisual effects, and physical memorial material~\cite{yi2021remember}. \textit{PetReunion} examines attitudes toward mixed-reality re-encounter, while related work explores 3D pet avatars in mixed reality~\cite{xiong2025petreunion_chi,lim2025avatars}. \textit{Bondi} uses a tangible device to support non-intrusive connection through touch, sound, and light~\cite{xiong2026bondi}. An affective-computing memorial system instead combines virtual pet behavior with multisensory hardware~\cite{wang2025memorialsystem}. Online pet-loss groups make disclosure, peer validation, and shared ritual central to the experience, while reflective game research shows how a symbolic world can organize bereavement reflection through metaphor~\cite{kim2026oscillation,lefevre2024loss}.

These modalities organize different relationships with remembrance. Installations and tangible objects embed memory in embodied ritual. Mixed-reality and avatar systems foreground re-encounter. Online groups make grief and response socially visible, while game worlds make reflection spatial and symbolic. They also carry different temporal commitments. An artifact may invite an occasional private ritual, an immersive representation may intensify a moment of presence, and a social platform can make continued participation visible. Kim and Park's account of \textit{compulsory grief} shows how that visibility can hold people in loss-oriented participation~\cite{kim2026oscillation}. The pet-specific systems reviewed here each foreground a primary form of encounter. Afterglow instead examines how private curation, human witnessing, lightweight world activity, and generated language in the Pet role can coexist within one environment. The resulting question concerns movement among these forms, including whether the memorial remains available through absence without demanding maintenance.

Animal--Computer Interaction foregrounds animal perspectives and agency in technologies intended for living animal users~\cite{mancini2011aci}. Afterglow instead studies caregiver-authored representations after an animal's death. Its visitor stance describes a mourner's relation to that representation rather than claiming participation or authorization by the animal.

\subsection{Generated Speaker Identity and Relational Authority}

Research on AI afterlives has primarily concerned human death and provides the closest account of what happens when preserved traces begin to speak in the first person. Scholarship on digital afterlives has examined consent, posthumous identity, dignity, commercialization, and the risk that fluency may be mistaken for relational continuity~\cite{ohman2017politicaleconomy,ohman2018ethical,jimenez2023griefbots}. Recent work follows generated personae through encoding, access, and dispossession and calls for meaningful transparency, mutual consent, and sensitive retirement procedures~\cite{lei2025aiafterlife,hollanek2024griefbots}. Morris and Brubaker map benefits and risks that include misrepresentation, dependency, and conflict~\cite{morris2025generativeghosts}. Spitale and Germani translate related concerns into inspectable constraints on consent, data sources, fidelity, disclosure, purpose, access, ownership, and agency~\cite{spitale2026digitalghosts}. Together, these accounts shift ethical attention from the apparent benevolence of a reply to the configuration that authorizes, circulates, and retires a generated persona.

Studies of grief-chatbot use show how relational meaning extends beyond source disclosure. Xygkou et al. found that mourners used chatbots as listeners, simulations of the deceased, friends, romantic partners, and emotion coaches. Some described deep relationships while recognizing the chatbot's artificial status, and most treated it as supplementary support~\cite{xygkou2023conversation}. Campbell et al. distinguish a thanabot relationship from continuation of the prior human relationship while allowing that it may support a continuing bond~\cite{campbell2025chatbots}. Stokes develops felt presence as a conceptual possibility for deathbot interaction~\cite{stokes2026sensing}. Manning et al. likewise found that participants crossed an intended boundary between third-person representation and first-person reincarnation while valuing affective resonance over factual fidelity~\cite{manning2026conversations}. Across these studies, artificiality can be recognized while generated speech still carries emotional weight.

Research outside pet bereavement also shows that the source of a response changes the support relation. Human responses in one distress-disclosure experiment were associated with warmth, whereas a responsive chatbot supported reappraisal through perceived competence~\cite{meng2023mediated}. Across nine experiments, Rubin et al. found that the same AI-generated empathic response was rated as more empathic and supportive when attributed to a human, and participants consistently preferred human interaction for emotional engagement~\cite{rubin2025empathy}. A preregistered two-week study outside bereavement similarly found that daily contact with a randomly assigned human peer reduced loneliness more than contact with a highly supportive chatbot~\cite{li2026humanpeer}. These findings distinguish AI acknowledgement from human reciprocity and motivate preserving the first opportunity for a person to respond.

Companion animals sharpen this authority problem. A person may leave premortem instructions, whereas a generated Pet persona is assembled entirely from caregiver-provided traces. It can feel intimate without acquiring authority to remember, forgive, or judge on the animal's behalf. The interface must therefore identify the speaker role as inferred and constrain what it may claim.

Agent-centered research clarifies the risks and lifecycle of the represented persona. Afterglow extends the unit of analysis to the surrounding memorial relation. Artifact-ecology research treats interactive artifacts as mutually shaping ensembles rather than isolated tools~\cite{bodker2011artifactecologies}. In a memorial ecology, that ensemble includes places, generated personae, human peers, data transitions, and platform rules. The design problem is therefore how these elements distribute intimacy and authority without returning every encounter to one simulated Pet.

\subsection{Relational Privacy and Shared Pet Memory}

Relational authority also depends on the materials from which a generated persona is made. Memorial data is relational because a photograph held by one caregiver may depict other people, a story may contain a family member's private history, and a pet profile may draw on memories that several caregivers understand differently. Prior work on post-mortem stewardship and socially produced memorial profiles shows that control over digital remains is distributed across relationships and among account holders~\cite{brubaker2013beyond,brubaker2014stewardship}. Pet memorials add a shared-caregiving asymmetry. Different caregivers may contribute or contest the materials from which the speaker identity is derived, extending authority beyond the uploader who assembled the profile.

Generation creates another boundary when contributed inputs become new outputs that circulate beyond their original context. Research on LLM conversation privacy shows that people treat conversational data as sensitive and hold context-dependent expectations about when and with whom it may be shared~\cite{tran2025privacy}. AI-afterlife work likewise connects preservation, access, transfer, and dispossession across a system lifecycle~\cite{lei2025aiafterlife}. We therefore treat privacy as a boundary across the whole memorial lifecycle. It begins with source and audience, continues through transformation, downstream use, and retention, and includes the ability to contest those uses or leave the system. These dimensions make it possible to ask who can see a memorial item and who can authorize what it becomes.

\subsection{From Visible Boundaries to Legible Restraint}

Boundaries around speaker roles and data become meaningful when people can recognize their scope and act on them. In emotionally sensitive interaction, a boundary often appears at refusal or breakdown, when a silent block or generic error can be interpreted as rejection. Seamful design offers another orientation by revealing boundaries so that people can form more accurate accounts of a system and act with its limitations~\cite{chalmers2003seamful}. Moderation research shows that explanations for removal decisions can shape how people understand a rule and participate afterward~\cite{jhaver2019moderation}. Human-centered responsible AI similarly turns ethical principles into actionable properties of interfaces and organizational practice~\cite{shneiderman2021responsibleai}. Trauma-informed computing adds that safety, trust, choice, and protection from retraumatization must shape model output, the surrounding system, and the research process~\cite{chen2022trauma}.

Disclosure and conversational behavior can pull in different directions. In one study, limitation disclaimers had no reported effect on trust, while an authoritative conversational style increased both trust and persuasiveness~\cite{metzger2024calibrated}. Research on AI companion loss and chatbot endings adds a temporal boundary. Shutdown can itself be experienced as loss, while agency and finality shape how people understand departure~\cite{banks2024deletion,poonsiriwong2026death}. Together, these studies motivate examining how a declared boundary carries through subsequent responses and changes in participation.

\section{Formative Survey and Initial Design Requirements}
\label{sec:formative}

\subsection{Research Orientation and Evidence Boundaries}

Afterglow was developed through an iterative RtD process in which survey responses, working prototypes, guided walkthroughs, and design deliberations informed one another. RtD knowledge remains answerable to the artifacts and situations from which it emerges~\cite{gaver2012expect}. We use this case to develop the three tensions and \textit{Legible Restraint} as situated, intermediate-level design knowledge. In H{\"o}{\"o}k and L{\"o}wgren's terms, a strong concept remains grounded in a particular artifact while offering a generative proposition that can travel to other design situations~\cite{hook2012strong}. Our empirical scope covers initial encounters, returning participants' encounters with a revision, and the resulting design interpretations. Therapeutic efficacy, grief reduction, clinical safety, and long-term behavior sit outside the present research questions.

The formative survey informed the initial prototype. Two rounds of prototype walkthroughs and roundtable discussion then supported its revision and formative evaluation~\cite{robinson2024formative}. The second round included separate roundtables for returning and newly recruited participants. Participants discussed the working prototype and responded to deliberately introduced scenarios involving reassurance, public acknowledgement, and moderation. From the outset, the team sought to distinguish the deceased animal from its generated representation and to limit the authority assigned to AI. Participants' responses could support that orientation or challenge how it shaped the experience. We examined those responses alongside changes to the prototype to develop the design argument.

\subsection{Ethics and Participant Safeguards}

The protocol was reviewed by the Institutional Review Board at Lehigh University and determined to be exempt. Recruitment focused on adults outside an immediate period of acute crisis. Participants provided informed consent, and the consent process presented the prototype as an experimental memorial system separate from therapy and crisis services. Walkthroughs were supervised, and participants could decline to discuss a loss or stop the session. Research outputs use project screenshots and aggregate text, with participant-created pet content kept private. The reviewed protocol adopted data minimization for these emotionally sensitive bereavement discussions. No audio or video was collected, reducing the retention of identifiable accounts of loss and grief. The interviewer's contemporaneous written notes constituted the sole session record. To protect participants' privacy, the full roundtable notes are retained privately. Appendix~\ref{app:coding-map} provides de-identified excerpts, the codebook, and links between evidence and design interpretations.

The survey and walkthrough materials were primarily in Chinese. Retained Chinese excerpts were first translated into English using Google Translate and then reviewed against the original wording by bilingual authors. Punctuation was normalized where needed. Quotations translated from Chinese are identified in the Findings and Appendix, and the Chinese wording remained in the note record. We report these accounts as culturally situated interpretations of grief practices and afterlife metaphors.

\subsection{Survey Instrument and Recruitment \texorpdfstring{($N=57$)}{(N=57)}}

We distributed a 24-item online survey in Chinese from September 20 to October 20, 2025, using snowball sampling in pet-owner and community networks. The first eight items recorded age, gender, race or ethnicity, current pet ownership, pet type, number of pets, prior pet loss, and time since loss. Four items then asked what having a pet meant to respondents, whether pet grief would be socially recognized, what support would be needed after loss, and how respondents preserved memories. The final 12 items introduced a possible virtual pet-memorial experience. They covered encounter format, initial interest, visual style, desired relationship with the pet, private or shared use, access to shared space, hoped-for benefits, AI moderation and disclosure, willingness to share, concerns, and desired features. Appendix~\ref{app:survey} provides the complete English translation, response formats, and options in their original order.

Three items were open-ended. The first asked, ``What does having a pet mean to you?'' The other two asked, ``What worries or concerns might you have about this kind of experience?'' and ``If you could try it, what features or functions would you want most?'' Each open-ended item received 51 responses. We distinguish the survey-level sample ($N=57$) from the item-level qualitative corpus ($n=51$ for each open prompt). All three exported open-item files retained the same 51 record numbers and submission timestamps. We refer to these matched survey records as S01--S51. The retained aggregate analysis recorded 46 of 57 respondents (81\%) as women. It also recorded that 32 of 57 (56\%) had experienced a companion animal's death. Among these 32 bereaved respondents, seven reported loss within the past year, 13 one to three years earlier, and 12 more than three years earlier.

The instrument functioned as a formative concept survey, eliciting reactions to virtual reunion, psychological support, AI-generated pet content, and AI moderation. Thirty-two respondents reported a companion-animal death, while the remaining respondents considered a possible future loss. We preserve this distinction between lived and anticipated perspectives in interpretation. Loss timing serves as descriptive context.

\subsection{Formative Analysis and Initial Requirements}

We used fixed-choice items descriptively and treated the 153 open-ended item responses as formative qualitative material. Table~\ref{tab:survey-needs} reports exact counts for the 32 bereaved respondents. Emotional comfort and preserving memories were prominent needs, while observation and simple interaction were selected alongside deeper exchange. Because relationship and benefit items allowed multiple selections, these choices describe compatible possibilities rather than exclusive user types.

\begin{table}
  \caption{Selected fixed-choice results among bereaved survey respondents ($n=32$). Counts are exact; percentages are rounded to whole numbers. Support, benefit, and relationship items allowed multiple selections. Style and medium rows list the two prototype-relevant alternatives from each item.}
  \label{tab:survey-needs}

\begin{tabularx}{\textwidth}{@{}p{0.19\textwidth}Y@{}}
    \toprule
    \textbf{Survey topic} & \textbf{Response counts} \\
    \midrule
    Support after loss & Emotional comfort 25/32 (78\%); remembrance 19/32 (59\%); practical aftercare help 16/32 (50\%); a community to share and talk with 14/32 (44\%) \\
    Hoped-for virtual benefits & Keeping or extending memories 24/32 (75\%); emotional comfort 23/32 (72\%); healing and psychological support 19/32 (59\%); talking with people with similar experiences 7/32 (22\%) \\
    Relationship with the pet & Observing a pet with its own life 21/32 (66\%); simple interaction 20/32 (63\%); deeper conversation or companionship 16/32 (50\%); activities or small games 11/32 (34\%) \\
    Visual style & As realistic as possible 21/32 (66\%); warm, healing cartoon style 9/32 (28\%) \\
    Encounter medium & Mobile app 15/32 (47\%); AR/VR 13/32 (41\%) \\
    \bottomrule
  \end{tabularx}
\end{table}

The preference for realism (21/32) set a demanding point of comparison for Afterglow's stylized world, with cartoon imagery preferred by 9/32. The prototype explored symbolic remembrance while carrying the demand for recognizable appearance and habits into the walkthrough inquiry. Participants could therefore question both the representation and the encounters built around it. Mobile access (15/32) and AR/VR (13/32) indicated interest in different encounter formats; the project pursued the mobile direction. Psychological support was a hoped-for benefit for 19/32 respondents, informing the inquiry into comfort and its boundaries rather than establishing a therapeutic outcome.

Time-since-loss groups provided descriptive context for variation in desired interaction. Deeper conversation or companionship was selected by 71\% of respondents within one year of loss ($n=7$), 69\% at one to three years ($n=13$), and 17\% beyond three years ($n=12$). In the last group, 75\% selected observer companionship. These small, cross-sectional groups motivated flexible interaction intensity without establishing a progression through grief stages.

For the qualitative reading, the first author compared matched S01--S51 records by question and across each respondent's three open responses. The analysis sought contrasting design directions rather than a second thematic model. Examples were selected when they made a repeated direction concrete or paired a desired experience with a concern in the same record. Some respondents wanted detailed visual and vocal likeness, personality, and dialogue. Others asked for quiet observation and warned that an animal suddenly speaking like a person would feel false. Respondents described potential support through memory preservation and acknowledgement while raising dependency, denial, renewed pain, and disappointment when generated likeness failed.

Desired-feature records gave quiet observation an everyday and temporal form. S01 imagined the pet as a companion ``living in a parallel world.'' S04 wanted to ``look quietly,'' while S07 and S23 wanted to see the pet happy and well in another world. S08 instead imagined ordinary care activities such as feeding, walking, and settling the pet to sleep. These responses framed the memorial as an inhabited place for observation, routine, and companionship as well as a site for addressing grief.

Paired concern and desired-feature responses made this ambivalence visible within individual records. S11 wanted to ``be able to talk with my pet'' while naming ``emotional dependence'' as the concern. S23 hoped to see the pet ``living well in another world,'' yet worried that moving between reality and fantasy could make it difficult to come to terms with the loss and prolong the pain. These written pairs illustrate anticipated value and anticipated risk coexisting within one account.

These responses and the literature informed three provisional requirements for the first prototype. The first supported recognizable and personally specific remembrance while framing generated content as caregiver-derived interpretation. The second combined acknowledgement and repeatable rituals with voluntary return. The third made the source and purpose of AI generation, sharing controls, and protective intervention understandable. The walkthrough study exposed conflicts within each starting requirement and motivated the tension families reported in Section~\ref{sec:findings}.

\section{Afterglow and RtD Iterations}
\label{sec:system}

\subsection{Distributing the Memorial Relationship Across Places}

\begin{figure*}
  \centering
  \begin{subfigure}[t]{0.235\textwidth}
    \includegraphics[width=\linewidth]{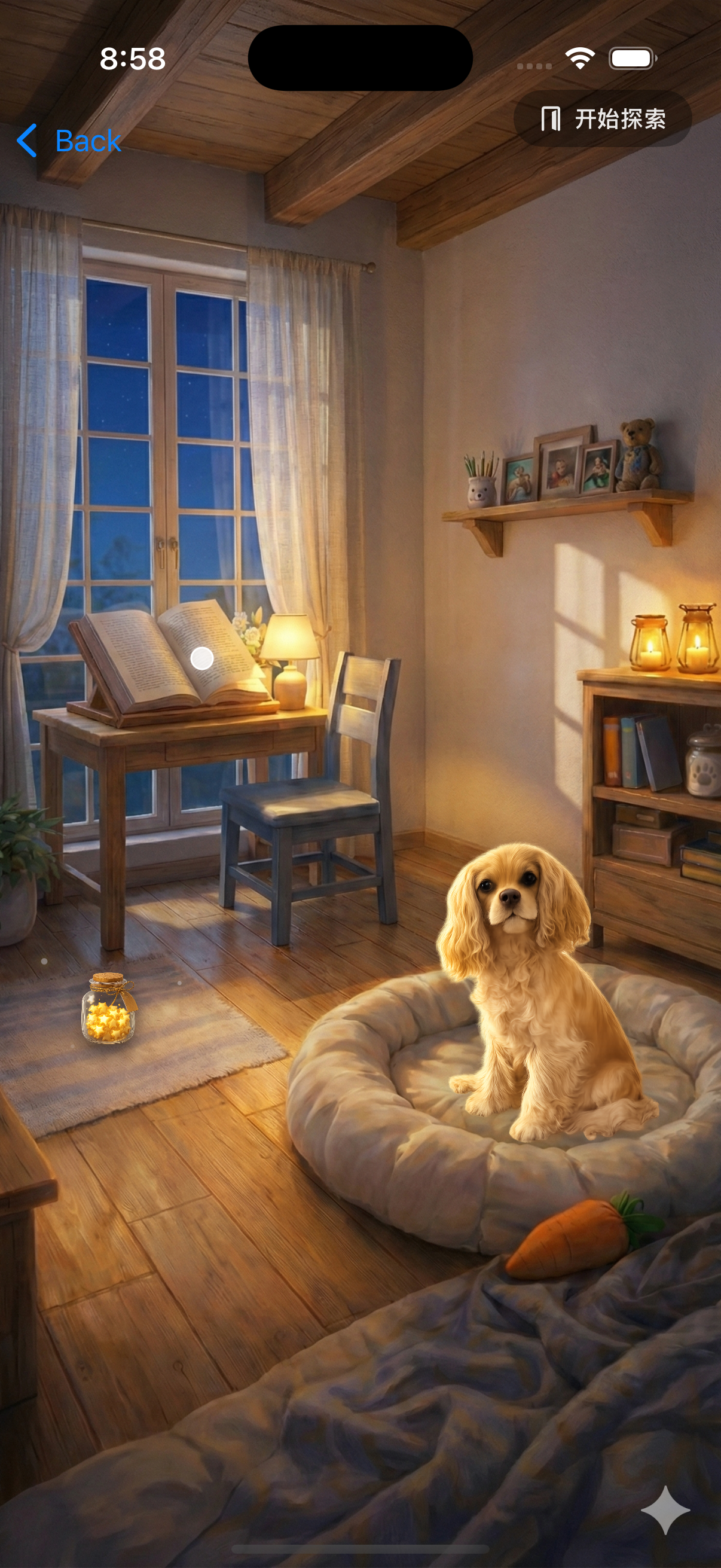}
    \caption{Memory Cottage}
  \end{subfigure}\hfill
  \begin{subfigure}[t]{0.235\textwidth}
    \includegraphics[width=\linewidth]{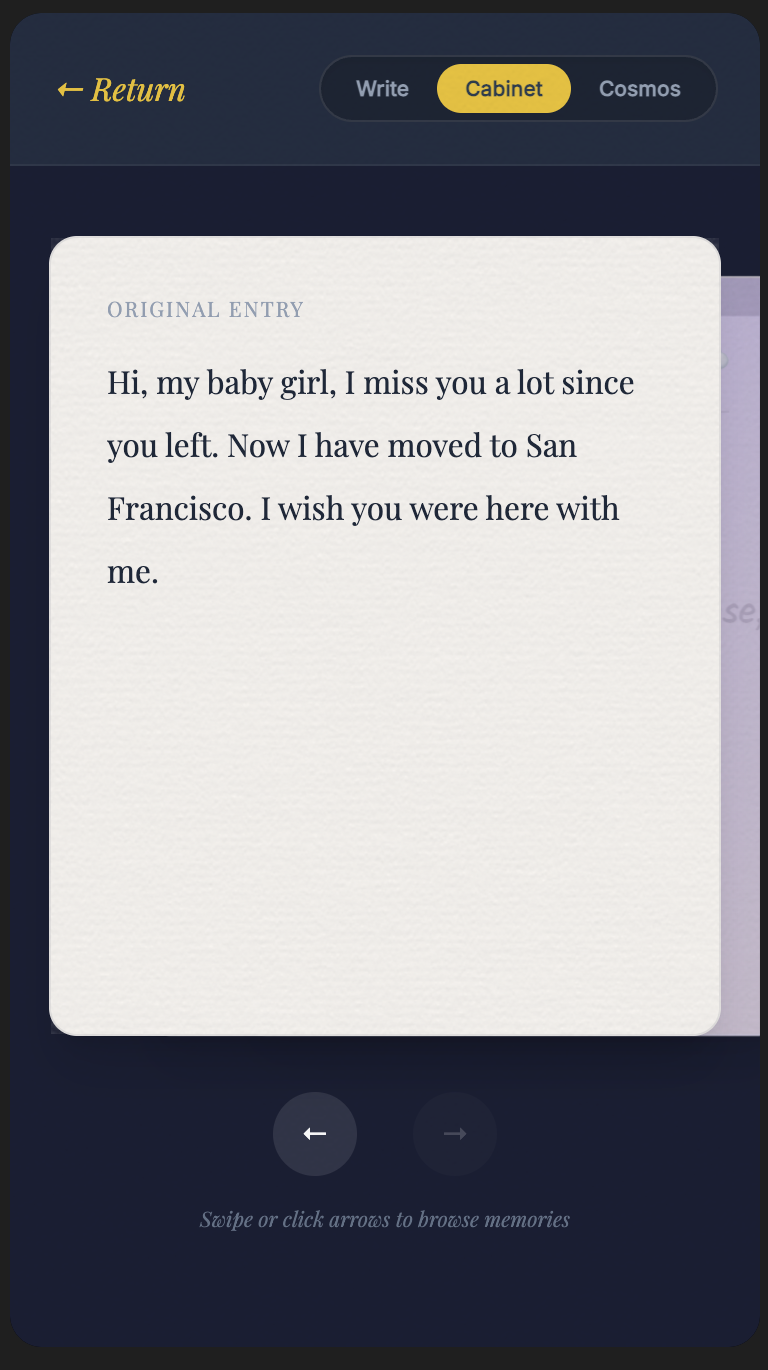}
    \caption{Memory journal}
  \end{subfigure}\hfill
  \begin{subfigure}[t]{0.235\textwidth}
    \includegraphics[width=\linewidth]{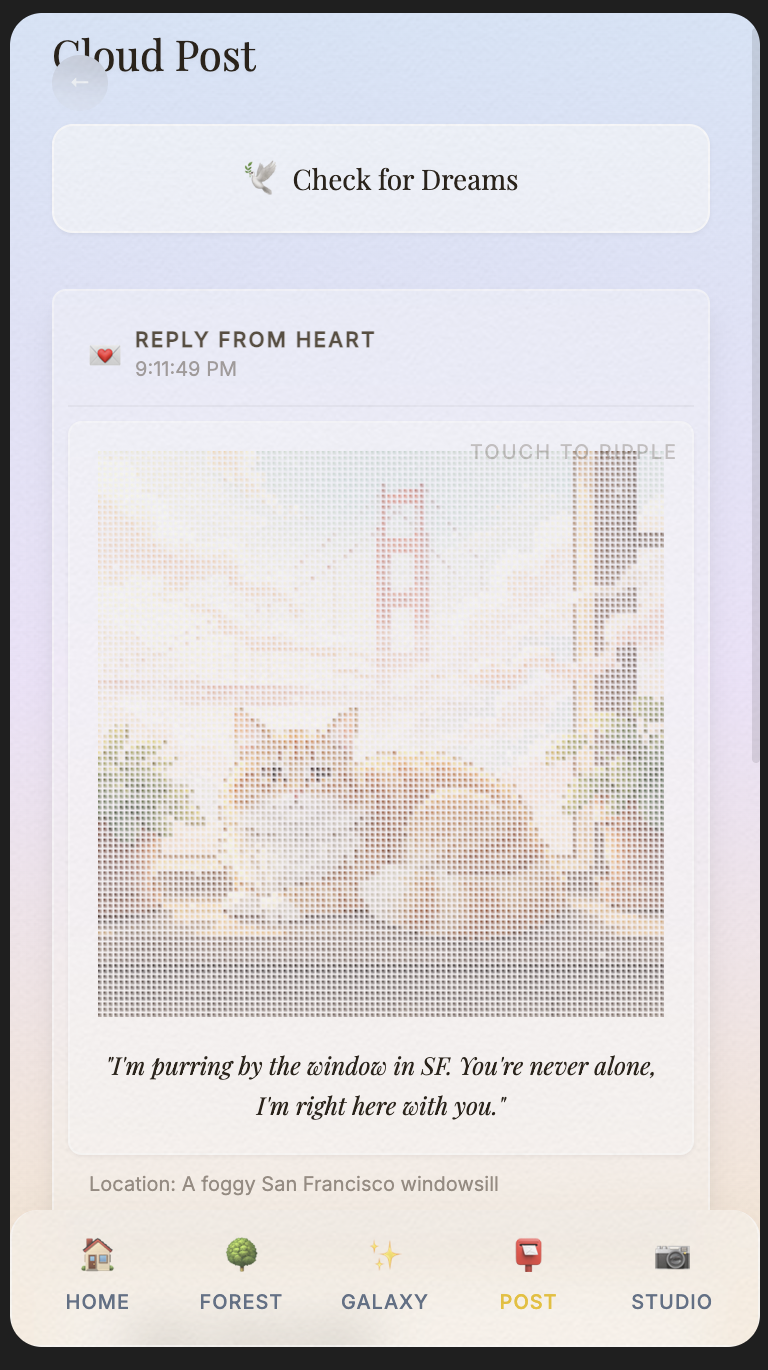}
    \caption{Generated reply}
  \end{subfigure}\hfill
  \begin{subfigure}[t]{0.235\textwidth}
    \includegraphics[width=\linewidth]{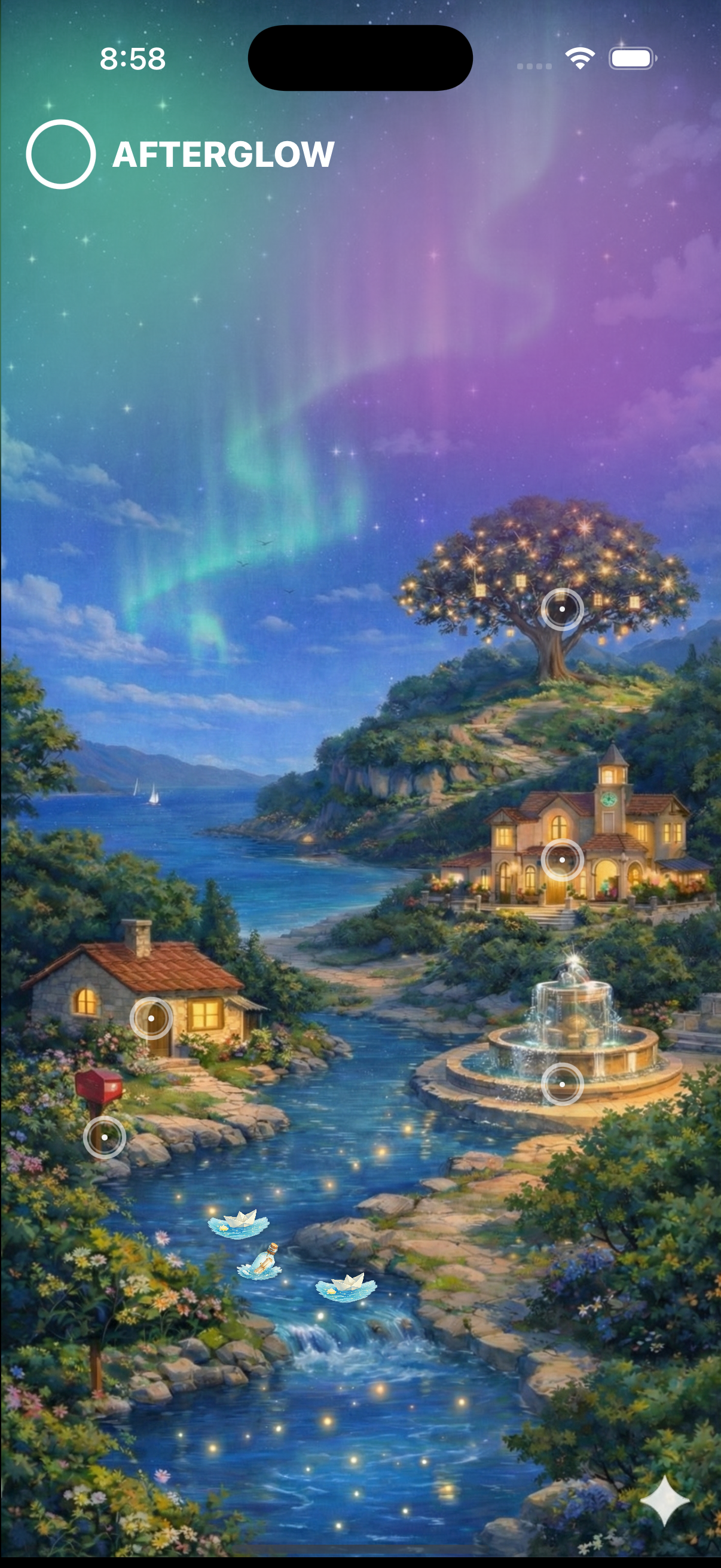}
    \caption{Shared world}
  \end{subfigure}
  \caption{Interfaces from the initial walkthrough build (\textsc{E1}). Participants in the first two roundtables encountered private remembrance, a generated reply, and a shared memorial world as parts of the same end-to-end journey.}
  \Description{Four vertical mobile screenshots show a cottage with a represented dog, a dark private memory-journal interface, a generated memorial reply, and a map-like shared memorial world with a river and interactive locations.}
  \label{fig:system}
\end{figure*}

Afterglow is a mobile memorial world organized around places and relational roles. The evaluated prototype included a Memory Cottage for private remembrance and a Memory River for anonymous or pseudonymous bottle messages. A Rainbow Post Office held pet-addressed letters, while a shared Forest let users visit traces of other relationships. The walkthrough began with the pet profile and private Cottage before introducing the River, Post Office, and Forest. After onboarding, these places remained independently accessible, supporting self-directed movement through the world. Figure~\ref{fig:system} shows retained interfaces from the evaluated-prototype period.

Figure~\ref{fig:system}c preserves one complete Postal output from the evaluated prototype, written in the pet's first-person voice. It reads, ``I'm purring by the window in SF. You're never alone, I'm right here with you.'' The sensory setting evokes a particular pet's life, while the reassurance comes from the represented animal rather than an identified platform narrator. In that voice, ``I'm right here with you'' claims continuing presence. The output shows how a recognizable memorial scene can acquire a relational claim through its speaker, even when the surrounding design treats the encounter as symbolic.

Each place gives the continuing bond a different social form. The Cottage supports quiet viewing and writing on the mourner's own terms. The River makes other mourners visible and permits optional human replies. The Post Office contains sparse, asynchronous Pet messages. The Forest represents relationships through trees, photographs, charms, visits, and small actions between users. The generated Pet therefore becomes one bounded encounter within a wider memorial world.

\paragraph{Game-inspired mechanics as social gestures.}
Water, tree growth, and collectible charms give the world visible change and provide low-threshold ways to interact with another person's memorial. Visiting a Forest and watering its tree was designed as a small acknowledgement when composing a consoling message felt difficult. The evaluated prototype tied some actions to visible resources and random rewards. The resulting design direction keeps visits and watering as optional traces of witness and companionship while separating resources, rewards, and memorial persistence from a user's absence.

\subsection{Role Distinctions Across Two Evaluated Builds}

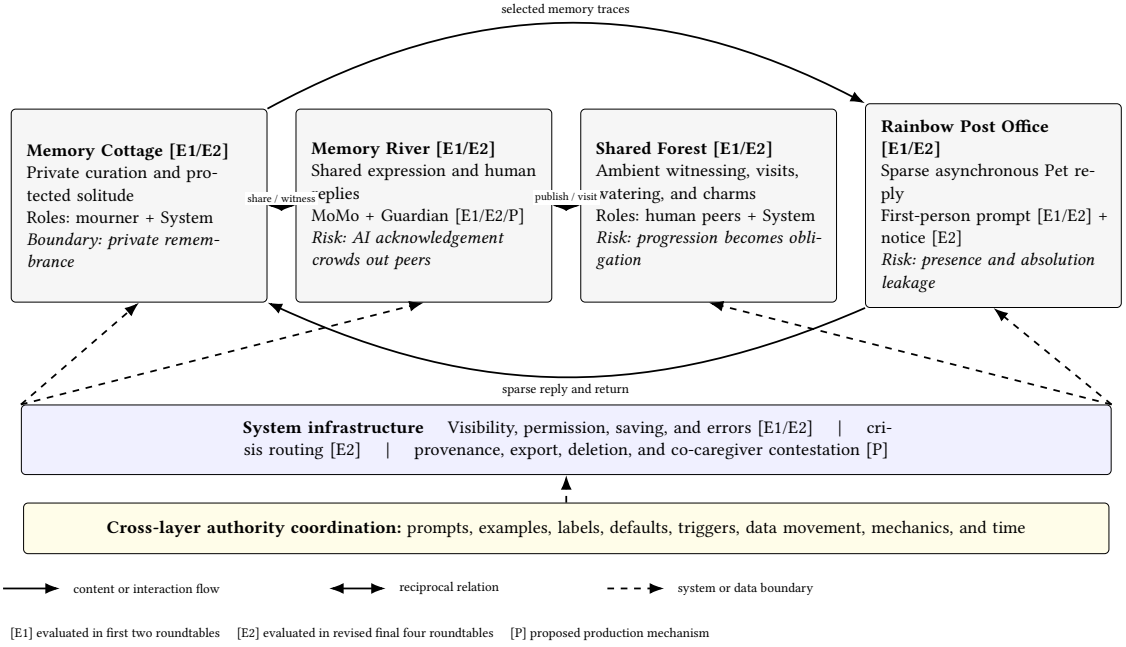
\begin{figure*}
  \centering
  \resizebox{0.98\textwidth}{!}{%
  \begin{tikzpicture}[
    >=Latex,
    place/.style={draw, rounded corners=2pt, fill=gray!7, align=left, text width=3.55cm, minimum height=3.05cm, inner sep=7pt, font=\small},
    infrastructure/.style={draw, rounded corners=2pt, fill=blue!6, align=center, text width=16.8cm, minimum height=1.1cm, inner sep=6pt, font=\small},
    coordination/.style={draw, rounded corners=2pt, fill=yellow!10, align=center, text width=16.8cm, minimum height=0.8cm, inner sep=5pt, font=\small},
    flow/.style={->, thick},
    reciprocal/.style={<->, thick},
    boundary/.style={->, dashed, thick}
  ]
    \node[place] (cottage) at (0,0) {\textbf{Memory Cottage \textsc{[E1/E2]}}\\Private curation and protected solitude\\Roles: mourner + System\\\textit{Boundary: private remembrance}};
    \node[place] (river) at (4.5,0) {\textbf{Memory River \textsc{[E1/E2]}}\\Shared expression and human replies\\MoMo + Guardian \textsc{[E1/E2/P]}\\\textit{Risk: AI acknowledgement crowds out peers}};
    \node[place] (forest) at (9.0,0) {\textbf{Shared Forest \textsc{[E1/E2]}}\\Ambient witnessing, visits, watering, and charms\\Roles: human peers + System\\\textit{Risk: progression becomes obligation}};
    \node[place] (postal) at (13.5,0) {\textbf{Rainbow Post Office \textsc{[E1/E2]}}\\Sparse asynchronous Pet reply\\First-person prompt \textsc{[E1/E2]} + notice \textsc{[E2]}\\\textit{Risk: presence and absolution leakage}};

    \node[infrastructure] (system) at (6.75,-3.7) {\textbf{System infrastructure} \quad Visibility, permission, saving, and errors \textsc{[E1/E2]} \quad $\vert$ \quad crisis routing \textsc{[E2]} \quad $\vert$ \quad provenance, export, deletion, and co-caregiver contestation \textsc{[P]}};
    \node[coordination] (coordination) at (6.75,-5.1) {\textbf{Cross-layer authority coordination:} prompts, examples, labels, defaults, triggers, data movement, mechanics, and time};

    \draw[reciprocal] (cottage) -- node[above, fill=white, inner sep=1pt, font=\tiny]{share / witness} (river);
    \draw[reciprocal] (river) -- node[above, fill=white, inner sep=1pt, font=\tiny]{publish / visit} (forest);
    \draw[flow] (cottage.north east) to[bend left=28] node[above, align=center, font=\scriptsize]{selected memory traces} (postal.north west);
    \draw[flow] (postal.south west) to[bend left=24] node[below, align=center, font=\scriptsize]{sparse reply and return} (cottage.south east);
    \draw[boundary] (system.north west) -- (cottage.south);
    \draw[boundary] (system.north west) -- (river.south);
    \draw[boundary] (system.north east) -- (forest.south);
    \draw[boundary] (system.north east) -- (postal.south);
    \draw[boundary] (coordination.north) -- (system.south);

    \draw[flow] (-2.15,-6.05) -- (-1.25,-6.05);
    \node[anchor=west, font=\scriptsize] at (-1.15,-6.05) {content or interaction flow};
    \draw[reciprocal] (3.0,-6.05) -- (3.9,-6.05);
    \node[anchor=west, font=\scriptsize] at (4.0,-6.05) {reciprocal relation};
    \draw[boundary] (7.4,-6.05) -- (8.3,-6.05);
    \node[anchor=west, font=\scriptsize] at (8.4,-6.05) {system or data boundary};
    \node[align=left, font=\scriptsize, anchor=north west] at (-2.15,-6.55) {\textsc{[E1]} evaluated in first two roundtables \quad \textsc{[E2]} evaluated in revised final four roundtables \quad \textsc{[P]} proposed production mechanism};
  \end{tikzpicture}%
  }
  \caption{Architecture and evidence status of Afterglow's memorial ecology. Four places separate private remembrance, shared expression, ambient witnessing, and pet-addressed exchange. The risk lines identify where AI acknowledgement, progression, or generated Pet speech can accumulate authority across that separation. \textsc{[E1]} marks the initial build evaluated in two roundtables, \textsc{[E2]} the revised final demo evaluated in four roundtables, and \textsc{[P]} a production proposal developed through the analysis.}
  \Description{A diagram connects four places: the private Memory Cottage, the public Memory River, the ambient Shared Forest, and the Rainbow Post Office. Solid arrows identify content and interaction paths, bidirectional arrows identify reciprocal relations, and dashed arrows identify System and data boundaries. Risk lines describe AI acknowledgement displacing peer response, progression becoming obligation, and generated pet speech leaking presence or absolution. A cross-layer coordination band lists prompts, examples, labels, defaults, triggers, data movement, mechanics, and time.}
  \label{fig:ecology-architecture}
\end{figure*}

Both walkthrough builds separated private remembrance, public witnessing, pet-addressed exchange, and platform intervention across places. The first two roundtables used the initial build (\textsc{E1}). Their discussion informed an iteration that made role boundaries more visible. The next four roundtables used the revised demo (\textsc{E2}), which identified MoMo and Guardian as AI and framed Postal output as symbolic memorial writing. Two \textsc{E2} sessions involved returning participants and two involved newly recruited participants (Section~\ref{sec:method}). Across both builds, prompts and examples still allowed generated roles to accumulate claims to continuing presence, lived experience, reassurance, and authority. The two rounds made these overlaps and the intervening revisions available for discussion. Figure~\ref{fig:ecology-architecture} maps the evaluated ecology and the production mechanisms developed from the analysis. Table~\ref{tab:roles} identifies the target boundary for each role.

\begin{table}
  \caption{Role distinctions used in the analysis and resulting design propositions. ``Target boundary'' states the design intention for each role.}
  \label{tab:roles}

\begin{tabularx}{\textwidth}{@{}p{0.12\textwidth}p{0.19\textwidth}YY@{}}
    \toprule
    \textbf{Speaker role} & \textbf{Where it appears} & \textbf{Permitted act} & \textbf{Target boundary} \\
    \midrule
    Pet & Rainbow Post Office letters and generated postcards & Short, sensory narrative grounded in remembered routines & Symbolic memorial narration; moral and clinical authority stays with people \\
    MoMo & Optional, delayed AI acknowledgement in shared spaces & Briefly recognize a post while identifying its artificial source & A human-first response window preserves the first opportunity for peer care \\
    Guardian & Before posting directed hostility in shared spaces & Explain the interpersonal boundary and invite revision & Targets interpersonal hostility while preserving intense grief expression \\
    System & Permissions, saving, privacy choices, and errors & State functional status and available next steps & Functional authority remains separate from the Pet's affection \\
    \bottomrule
  \end{tabularx}
\end{table}

\paragraph{Pet as a narrative device.}
The target Pet role uses short, concrete, sensory language and reserves diagnosis, life advice, euthanasia judgment, and moral absolution for people with relevant expertise or relational standing. The retained backend combined this target with a Postal prompt that required pet-like first-person language from Rainbow Planet and implied continuing nearness through wind, starlight, or warmth. This mismatch exposed prompt-level role leakage. First-person framing and sampled examples could confer authority that a later boundary statement attempted to withdraw. It redirected redesign across the system prompt, examples, interface labels, and generated output, making restraint an architectural concern.

\paragraph{MoMo as acknowledgement.}
MoMo was discussed as an AI-generated acknowledgement for shared expression. The initial River surface presented its reply within the shared space. The revised demo added a persistent AI label and an off switch. The retained path could reply automatically before the status of human response became known. The proposed human-first window begins when a public memory appears. During a disclosed waiting interval, the interface reserves response for people. If no human reply arrives, the author may explicitly request one short MoMo acknowledgement. A human reply keeps MoMo off by default. This sequence gives human reciprocity the first opportunity while preserving an explicitly requested backstop that identifies its artificial source.

\paragraph{Guardian as boundary keeper.}
Guardian was staged around interpersonal harm in proposed shared content. This included hate speech, threats, malicious ridicule, and harassment. The intended response explains the shared-space rule, invites revision, and preserves grief intensity as legitimate expression. The retained gate combines keyword lists with a binary LLM judgment of attacks, insults, threats, mockery, and hostile provocation. Its output schema contains no field for the target of an expression, so self-blame and interpersonal attack enter the same decision path. Guardian therefore moderates interpersonal conduct rather than certifying the truth or emotional safety of a Pet message. The revised demo identified \textit{Guardian (AI)} visibly and separated crisis routing from public-content moderation.

\paragraph{System as infrastructure.}
Functional messages use plain infrastructural language under the System role. Upload failures, privacy choices, and missing fields therefore retain an infrastructural source. This separation keeps functional authority distinct from Pet affection and makes refusal sources easier to distinguish.

\begin{figure}
  \centering
  \includegraphics[width=0.30\textwidth]{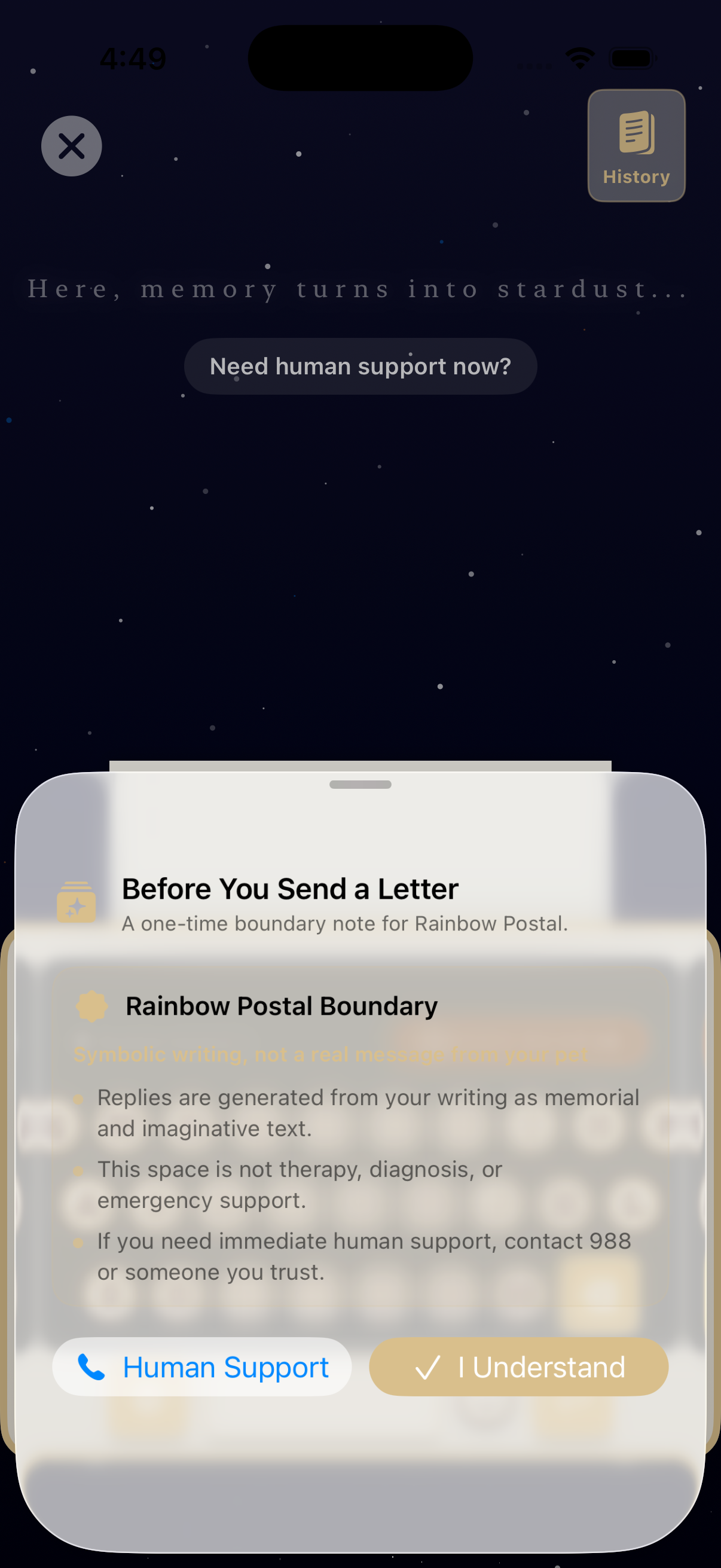}
  \caption{Postal generation notice in the revised demo interface (\textsc{E2}). It frames letters as symbolic memorial writing generated from caregiver-provided traces. This implementation screenshot documents the disclosed source and status of Pet speech.}
  \Description{A mobile Postal screen explains that generated letters are symbolic memorial writing rather than real messages from the pet or therapy.}
  \label{fig:poststudy-boundaries}
\end{figure}

\subsection{Privacy and Sharing Boundaries}

Afterglow's intended privacy boundary is place- and role-specific and appears at relevant transitions. The Cottage and selected Pet letters support private remembrance. The River asks users to choose whether a contribution is private or public. Public contributions become eligible for community visibility, optional MoMo acknowledgement, and Guardian review. One retained composer still defaults to public, making a private-first default a concrete production requirement rather than an accomplished boundary. The System role presents visibility state and AI involvement. In Legible Restraint terms, a sharing control should communicate its source (the user or platform rule), scope (which artifact and audience), and consequence (storage, model processing, generated response, or publication).

Audience control forms one layer of relational privacy. A user may have permission to upload a photograph while another caregiver disputes its use in generated speech. A private input may produce a postcard that is later shared, and deletion of the visible artifact provides limited evidence about copies in models, logs, or backups. We therefore treat consent to contribute, consent to generate, and consent to share as separable decisions~\cite{brubaker2014stewardship,lei2025aiafterlife,tran2025privacy}. A complete system would also need provenance for contributed materials, co-caregiver objection and removal paths, retention disclosure, and contestable permissions for shared memories.

The revised final demo partially realizes this account. It stores app state locally and uses a locally hosted Ollama model by default. Public River text is sent to the local backend for moderation before release, and safety logs retain structured metadata in place of raw text. These choices reduce third-party exposure in the research configuration. Production development still requires a private-first default across composers, co-caregiver dispute workflow, provenance ledger, verified deletion across providers, and tested access control. Posting, reacting, and replying can also add water to the Memory Tree, coupling a visible social gesture with memorial activity.

\subsection{Temporal Openness and Designing for Goodbye}

Afterglow's temporal question is whether a memorial world can invite return while tolerating dormancy. Its locations remain independently accessible, visits can be brief, and access persists through absence. Water, tree growth, and collection direct attention toward memories and peer interaction, but they also make activity measurable. Repeated use may signify care, recognition, progress, or obligation. Pause, export, close-this-memory, and return-later paths emerged as later design directions.

Non-use can be a meaningful relation to technology, while involuntary deletion or shutdown of an AI companion can be experienced as another loss~\cite{satchell2009beyond,banks2024deletion}. Research on chatbot endings further shows how agency, finality, and user initiation shape the experience of departure~\cite{poonsiriwong2026death}. A person may return for years, use the system briefly, export selected memories, or delete the app. The platform should treat each trajectory as a valid relation to remembrance, supporting user-defined timelines for continuing bonds~\cite{paumen2026neversaygoodbye}. Peer acknowledgement, felt social support, and intact dormancy become primary criteria for world mechanics. Engagement metrics serve as diagnostic evidence within this account.

\subsection{Implementation and Scope}

The iOS prototype is implemented in SwiftUI with a Node.js/Express backend. The retained configuration serves \texttt{gemma3:12b-it-q4\_K\_M} through local Ollama. Postal and MoMo each sample four local few-shot examples per request. Postal uses temperature 0.8, top-$p$ 0.9, and a 220-token cap. MoMo uses temperature 0.6, top-$p$ 0.9, and a 100-token cap. The moderation classifier uses temperature 0, top-$p$ 0.2, and a six-token cap to return a binary judgment. System messages are deterministic interface strings, distinct from the three generative roles. Asynchronous jobs delay selected Pet replies and frame the exchange as sparse and asynchronous. Appendix~\ref{app:prompts} summarizes the retained role instructions. The supplementary prompt listing provides the full templates and parameters.

The prototype functions as a design probe. Probabilistic prompt constraints, binary moderation ambiguity, and users' agent attributions made the distribution of authority a central RtD concern. The first-person Postal prompt positioned continuing pet presence more strongly than the analytic target boundary. This internal tension brought role separation into the inquiry itself. The study traced how a provisional artifact distributed and contested authority.

\section{Roundtable Interview and Prototype Walkthrough Study}
\label{sec:method}

\subsection{Participants and Roundtable Procedure \texorpdfstring{($N=20$)}{(N=20)}}

We conducted six 60-minute online facilitated roundtables from November 26, 2025 to February 19, 2026, combining discussion with prototype walkthroughs. Round 1 comprised two sessions with ten participants using the initial build (\textsc{E1}), held between November 26 and December 20, 2025. We revised the presentation of AI roles, generated content, crisis routing, and optional world mechanics in response to these discussions. Round 2 comprised four sessions using the revised demo (\textsc{E2}), held from January 17 to February 19, 2026. All ten Round 1 participants returned in two sessions; ten additional participants were newly recruited for the other two sessions. The study therefore included 20 unique participants and 30 participant attendances across both rounds (Table~\ref{tab:roundtable-cohorts}).

\begin{table}
  \caption{Participation across the two rounds of formative evaluation. Returning and newly recruited participants attended separate Round 2 sessions.}
  \label{tab:roundtable-cohorts}

\begin{tabularx}{\textwidth}{@{}p{0.13\textwidth}p{0.14\textwidth}p{0.17\textwidth}Y@{}}
    \toprule
    \textbf{Build} & \textbf{Sessions} & \textbf{Participants} & \textbf{Prior encounter with Afterglow} \\
    \midrule
    \textsc{E1} & 2 & 10 & Initial walkthrough \\
    \textsc{E2} & 2 & 10 returning & All participants from \textsc{E1} returned to encounter the revision \\
    \textsc{E2} & 2 & 10 new & First walkthrough, using the revised build \\
    \bottomrule
  \end{tabularx}
\end{table}

This arrangement brought revisiting and first-contact perspectives into the revised-build evaluation. The returning cohort could encounter changes in light of the earlier prototype; the new cohort encountered \textsc{E2} without that prior walkthrough. We analyze discussion-level tensions across this iterative process, rather than estimating a controlled effect of build version or a within-person change score.

Six of the 20 unique participants had completed the formative survey and volunteered for follow-up. The other 14 were recruited separately through pet-owner and community networks. These connected formative samples therefore partially overlap. The first author facilitated every session, and participants had no prior personal relationship with the facilitator. Sessions moved between Chinese and English according to participant preference. Participants were 18--35 years old; three identified as men and 17 as women, and all identified as Asian. Education information was available for 18 participants, comprising two undergraduate students, six bachelor's degree holders, and ten graduate degree holders. Two participants left education unreported. We report these characteristics in aggregate to reduce identifiability in small-group bereavement discussions. All participants reported companion-animal loss, with seven within the previous year, eight one to three years earlier, and five more than three years earlier.

Each session followed the same semi-structured sequence. The facilitator first introduced the project and demonstrated Afterglow as an end-to-end memorial journey. The walkthrough covered private curation in the Cottage, shared expression and peer witnessing in the River and Forest, pet-addressed exchange in the Post Office, and platform boundary encounters involving MoMo, Guardian, and System. Participants then described their initial feelings and reactions. The remainder of the session used open group discussion, allowing participants to revisit any place or function and respond to one another's interpretations. Near the close of the Round 2 sessions, the facilitator asked when participants could imagine first using Afterglow and for how long. The reported timing tallies draw on all 20 participants' Round 2 answers (FN-10). The group format made a comment or discussion episode the primary analytic unit. The supplementary coding map reproduces the complete facilitation sequence. Loss timing provides descriptive context.

The facilitator recorded notes during each session and expanded their context immediately afterward. The note record distinguished statements captured in direct-speech form from summaries of longer exchanges. A speaker link was considered reliable when a direct-speech entry was saved with that participant's session identifier at the time of the comment. We use quotation marks only for these speaker-linked entries, reported as note-based quotations with de-identified labels I01--I20. These quotations preserve the wording available in the written record. Discussion episodes retained only in summary form are reported as analytic paraphrases with FN labels.

Within the open discussion, the facilitator introduced the same three focused design probes to make system authority consequential. They covered a Pet response that absolved a caregiver's euthanasia decision, an AI acknowledgement under a public post, and intervention against directed hostility. The probes sharpened key moments in the broader journey and elicited interpretations of speaker identity and authority. The interviewer notes preserve the scenario topics and positions that emerged. FN-07 records a response to the staged Pet-absolution scenario, while FN-10 records answers to the closing timing questions. For other selected excerpts, elicitation status remains uncertain because spontaneous remarks and responses to facilitator prompts were inconsistently distinguished. Our interpretation therefore concerns tensions made salient within the guided discussion; spontaneous recurrence remains unestablished.

Both rounds used the same end-to-end journey through the four places and role scenarios represented in Figure~\ref{fig:system}. The intervening iteration added the persistent \textit{MoMo (AI)} and \textit{Guardian (AI)} labels, Postal generation notice, clearer crisis routing, and non-decaying optional mechanics while preserving the connected journey. After the \textsc{E2} sessions, implementation work was limited to refining the generated-photo feature. Participants had encountered that feature and its memorial scenario, while the later photo outputs serve as implementation documentation.

\subsection{Designer-Led Interpretive Analysis and Coding Traceability}

\begin{figure*}
  \centering
  \resizebox{0.96\textwidth}{!}{%
  \begin{tikzpicture}[
    >=Latex,
    source/.style={draw, rounded corners=2pt, fill=gray!10, align=center, text width=16.5cm, minimum height=1.15cm, font=\small},
    process/.style={draw, rounded corners=2pt, fill=blue!5, align=center, text width=12.5cm, minimum height=1.05cm, font=\small},
    family/.style={draw, rounded corners=2pt, fill=green!8, align=center, text width=5cm, minimum height=2.0cm, inner sep=6pt, font=\small},
    synthesis/.style={draw, very thick, rounded corners=2pt, fill=yellow!12, align=center, text width=12cm, minimum height=1.15cm, font=\small\bfseries},
    excluded/.style={draw, dashed, rounded corners=2pt, fill=gray!5, align=center, text width=3.7cm, minimum height=1.15cm, font=\small}
  ]
    \node[source] (notes) at (0,0) {Contemporaneous interviewer notes from facilitated roundtables\\Written-note record; selected key statements; designer-led coding};
    \node[process] (coding) at (0,-1.8) {Comment or discussion episode $\rightarrow$ descriptive/open code $\rightarrow$ focused code};

    \node[family] (connection) at (-6,-4.2) {\textbf{Connection and witnessing}\\10 focused codes\\[3pt]\textbf{Returnable connection vs. emotional obligation}};
    \node[family] (representation) at (0,-4.2) {\textbf{Representation}\\8 focused codes\\[3pt]\textbf{Recognizable likeness vs. ontological clarity}};
    \node[family] (authority) at (6,-4.2) {\textbf{Authority and safety}\\5 focused codes\\[3pt]\textbf{Protective intervention vs. surveillant authority}};

    \node[synthesis] (synthesis) at (0,-7.0) {Design implications $\rightarrow$ exploratory synthesis: Legible Restraint};
    \node[excluded] (stage) at (8.4,-7.0) {Stage-Sensitive Needs\\Provisional code};

    \draw[->] (notes) -- (coding);
    \draw[->] (coding.south west) -- (connection.north);
    \draw[->] (coding.south) -- (representation.north);
    \draw[->] (coding.south east) -- (authority.north);
    \draw[->] (connection) -- (synthesis.north west);
    \draw[->] (representation) -- (synthesis.north);
    \draw[->] (authority) -- (synthesis.north east);
    \draw[->, dashed] (coding.east) .. controls (8.5,-1.8) and (9,-5.5) .. (stage.north);
  \end{tikzpicture}%
  }
  \caption{Roundtable analysis logic. The interviewer coded contemporaneous notes at the level of comments and discussion episodes, organized 23 reported focused codes into three tension families, and retained one additional code as provisional. Appendix~\ref{app:coding-map} provides definitions for all 24 codes and documents analytic traceability.}
  \Description{A flow diagram begins with contemporaneous written interviewer notes analyzed through designer-led coding. Comment and discussion episodes move through descriptive and focused coding into three tension families. These inform design implications and Legible Restraint. A dashed branch marks stage-sensitive needs as provisional.}
  \label{fig:coding-map}
\end{figure*}
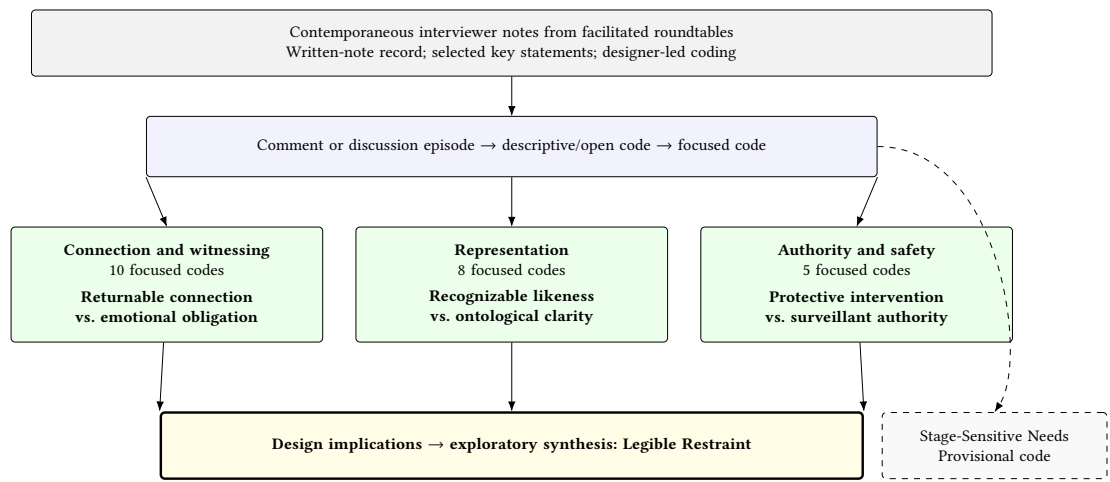

We used designer-led interpretive thematic analysis~\cite{braun2006using,zimmerman2007research,gaver2012expect}. The first author occupied a designer-researcher position, developing the prototype, introducing its journey, facilitating every session, expanding the contemporaneous notes, and conducting the coding. They entered the study with explicit commitments to ontological clarity and a non-clinical memorial framing. This situated role supported artifact-specific interpretation of prompts, interface transitions, and prototype versions. The analysis made that position inspectable by distinguishing speaker-linked wording from analytic paraphrase, retaining counterpositions, and checking code-to-design links against versioned artifacts. The 153 open-ended survey responses formed the separate formative analysis in Section~3, while the roundtable coding map draws directly from contemporaneous discussion notes. For the roundtables, the interviewer-designer coded notes at the level of comments and discussion episodes. The record retains the progression from descriptive/open codes to focused codes, tension families, and design implications. De-identified code labels and aggregated themes entered the design record and manuscript. The focused codes covered socially minimized grief, protected expression, symbolic bonds, recreation, and AI identity, together with moral reassurance, refusal, role clarity, and emotional containment. Figure~\ref{fig:coding-map} summarizes this analytic logic, while Appendix~\ref{app:coding-map} provides the complete focused codebook.

Our analysis interpreted design tensions across selected discussion episodes from the RtD sessions. Theme construction emphasized design consequence rather than prevalence, retaining a counterposition when it exposed conflict in an interface, prompt, transition, or role boundary. Desires for reassurance remained alongside doubts about machine judgment; requests for acknowledgement remained alongside the value of solitude and human witnessing. These counterpositions challenged any simple reading of restraint as participants' preferred endpoint. The three families synthesize such conflicts rather than ranking attitudes toward AI or estimating how many sessions endorsed each position. \textit{Stage-Sensitive Needs} remains provisional and outside the three reported tension families. The selected speaker-linked excerpts in the Findings represent 12 participants. Other evidence remains linked to discussion episodes rather than individuals. The codebook and artifact history make the resulting situated RtD interpretation traceable.

\subsection{Traceability from Evidence to Design}

Table~\ref{tab:traceability} connects the retained evidence and walkthrough interpretations to design responses, recording their rationale and provenance.

\begin{table}
  \caption{Traceability from the three tension families to design responses. \textsc{E1} denotes the initial build evaluated in two roundtables, \textsc{E2} the revised final demo evaluated in four roundtables, and \textsc{P} proposed production mechanisms.}
  \label{tab:traceability}

\begin{tabularx}{\textwidth}{@{}p{0.19\textwidth}YYY@{}}
    \toprule
    \textbf{Tension family} & \textbf{Design requirement} & \textbf{Afterglow response} & \textbf{Residual risk} \\
    \midrule
    Returnable connection vs. emotional obligation & Preserve private remembrance while making human witnessing and acknowledgement available & \textsc{E1}: Cottage solitude, River witnessing, delayed Pet messages, and watering. \textsc{E2}: distinct MoMo labeling and optional, non-decaying mechanics. \textsc{P}: opt-in acknowledgement after a human-first window & Automatic AI acknowledgement can displace human reciprocity; watering and collection can create maintenance debt; one sharing surface defaults to public \\
    Recognizable likeness vs. ontological clarity & Support recognition while revealing the source, transformation, and audience of generated material & \textsc{E1}: stylized imagery and first-person Postal exchange. \textsc{E2}: generation notice and visible role separation. \textsc{P}: bounded prompt redesign and shared-source dispute paths & The retained first-person Postal prompt implied ongoing pet presence; shared-source disputes remain unresolved \\
    Protective intervention vs. surveillant authority & Limit moderation to proposed shared content, preserve intense personal grief, and identify who sets the boundary & \textsc{E1}: staged directed-hostility scenarios. \textsc{E2}: visible Guardian labels and separate crisis routing. \textsc{P}: target-sensitive moderation & The binary gate uses a hostility decision with target and cultural context left unspecified \\
    \bottomrule
  \end{tabularx}
\end{table}

\section{Findings}
\label{sec:findings}

Our interpretive analysis identified three design tensions that became salient during the prototype walkthroughs and roundtable discussions. Each family brings together desires that the artifact made difficult to satisfy at once. The selected quotations draw from six, five, and three participants across the three themes, respectively, with some appearing in more than one theme. These figures describe quotation coverage only. Analytic paraphrases add selected discussion episodes, while the timing tally describes anticipated entry and duration. Appendix Table~\ref{tab:fieldnote-exemplars} links each retained excerpt to focused codes and design implications.

\subsection{Theme 1. Returnable Connection versus Emotional Obligation}

Roundtable discussion held two needs together. Pet grief merits social recognition, while disclosure, reply, and return remain voluntary. I14 described the social difficulty directly, ``I can't tell my coworkers I'm still crying over a cat. Here, it feels allowed.'' Seeing that other people also mourn companion animals gave the River and shared Forest value before anyone began a direct conversation. I09 explained why that human presence mattered, ``Seeing that someone else is missing their dog today... that helps more than a bot telling me it's okay.''

Protected solitude remained equally important. Participants could write or revisit memories privately in the Cottage, while River contributions made audience and response consequential. I15 described expression as a way of living with loss, ``Writing this is not necessarily about seeking sympathy; it is about making the loss easier for me to live with'' (translated from Chinese). Public expression created a different vulnerability. I06 described the cost of silence, ``If I put something out and nothing happens, it feels worse. Even just a small reply makes it feel like someone noticed.'' Read alongside I09's account, I06's concern makes acknowledgement and human witnessing complementary possibilities. It supports considering a brief MoMo acknowledgement when human response is absent. These accounts raised a design question about who should acknowledge a post and when.

The world also offered watering, tree growth, reactions, and collection as lightweight routes into interaction. Their design rationale was social. A person could acknowledge another memorial through a small gesture before composing a consoling message. Visible growth and accumulation, however, can also make activity look like progression and connect remembrance to a record of participation. The central question is whether that activity signifies recognition or care that must be performed.

Other discussion episodes gave lightweight activity another meaning. Participants wanted the app to feel lived-in rather than resemble an online cemetery (FN-05). They imagined opening it at night or while alone, including long after the pet's death, to share recent pleasures and difficulties with the pet (FN-06). I01 described a likely rhythm, ``I would probably open this quite often, but I wouldn't stay inside for very long. Maybe just check if there's a new message, water the tree, and then close it.'' Small activities offered a reason to return for private company, observe the represented pet appearing happy in its own place, and encounter something beyond repeated grief talk. Participants described this as companionship in another form. World mechanics could therefore sustain everyday return as well as mediate peer acknowledgement.

Closing questions in Round 2 made the anticipated trajectories more concrete (FN-10). Five of 20 participants imagined beginning to use Afterglow soon after the pet's death. Ten preferred to wait one to three months, when the loss might feel less acute. Three were unsure that they would create an account but could imagine returning to read what others had written, while two expressed doubts about their readiness for such a system. When asked about duration, 15 imagined using it for several months to one year. Six within this group anticipated writing daily at first. Five participants were uncertain about duration, and participants noted that the trajectory would depend on the circumstances of the pet's death. Across these answers, access remained more important than a fixed schedule. Participants wanted to enter during fragmented moments to write or speak, even after regular use had tapered off. These counts describe anticipated use rather than observed longitudinal behavior.

Another discussion episode drew a sharper boundary between visiting and caretaking (FN-09). Requiring repeated feeding or petting, as with a virtual pet, would make the mourner feel responsible for controlling the represented animal's life and guilty during periods of absence. I13 put this concern in temporal terms, ``I don't want to have to feed or pet them like a virtual pet. If I became busy and stopped coming, I would feel guilty, as though I had neglected them again. I may not need the app forever, but if I return one day, I still want to find my dog happy here'' (translated from Chinese). The desired world therefore allowed the person to step away and still return to a Pet who remained well.

Returnable connection therefore frames game structure as an architectural tension. Private remembrance, ambient human presence, direct peer response, AI acknowledgement, and symbolic Pet messages each create a different reason to return. The connection becomes obligatory when the platform equates visible activity or reciprocal response with the adequacy of care.

\subsection{Theme 2. Recognizable Likeness versus Ontological Clarity}

Desired likeness centered on specificity. Survey and walkthrough records asked for a pet who felt particular through appearance, vocal likeness, habits, personality, and recognizable routines. The same records preserved discomfort with inaccurate likeness, uncanny realism, or adult-like dialogue that made the system feel like a chatbot using the pet as a mask. I04 located the rupture precisely, ``If the cat suddenly starts giving me life advice, it feels like a chatbot pretending to be my cat, and that breaks the illusion.''

The survey's quiet-observation accounts carried into the roundtable discussion as a visitor stance. Participants imagined the represented pet having activity ``over there'', with the person entering as a visitor and observer. I02 explained, ``I don't want to be telling them what to do all the time. It feels nicer if they invite me into something, like `today I want to show you this place.','' I03 connected that stance to release, ``I like the idea that they've gone to a new world and they're free now. I don't want to keep ordering them around with tasks here. I just want to see that they're okay.'' This imagined autonomy shifted the world away from pet management, although it could also strengthen the fiction that the pet continued to live elsewhere.

Pet-initiated postcards and small environmental changes supported this visitor stance. Symbolic representation was also associated with imagined memorial images. I01 described one such absence, ``When you have a childhood pet, adult-you never had a chance to take a proper photo together. This lets you fill in that missing picture that you always wished you had.'' I11 drew a spatial boundary around another wished-for scene, ``If I saw my dog in my real surroundings through mixed reality, it would feel a little creepy. But I would still like a place where the dog I lost could meet the dog who is still alive, as if they were playing together at home again'' (translated from Chinese). The contrast placed comfort in a bounded memorial encounter rather than an overlay on everyday surroundings. In both examples, value came from staging a personally specific, wished-for scene.

The survey's preference for realistic appearance and the walkthrough concerns about artificial authority address different aspects of representation. Likeness can support recognition, while life advice or claims to continuing presence can change what the Pet appears entitled to say. Ontological clarity concerns how vivid, personally specific representation remains compatible with a bounded speaker role.

\subsection{Theme 3. Protective Intervention versus Surveillant Authority}

Protection changed meaning depending on who was protected, from what, and by which role. The coding records reserved an AI Pet for bounded expression and preserved intense grief as legitimate. Functions such as organizing material, generating optional symbolic images, or producing short sensory messages fit the Pet role, while life advice and morally consequential judgment shifted authority toward a therapist-like agent. I05 accepted visible imperfection as part of that boundary, ``I'm fine if it's not perfect, as long as it doesn't try to act like a real therapist. I'd rather it be a bit clumsy and obviously my pet.'' Adult-like advice also exposed the model behind the pet representation.

Euthanasia and guilt made this boundary especially visible. In the staged Pet-absolution discussion, wanting reassurance coexisted with concern about an unqualified ``you did the right thing'' spoken by the Pet (FN-07). This analytic paraphrase preserves both the desire for comfort and the sense that machine-delivered moral certainty felt hollow. The design issue is how to acknowledge the request while keeping morally consequential judgment with people who have relevant standing.

The same records preserve conditional support for Guardian intervention against harm to others and resistance to a moderator that appeared to monitor grief itself. I07 drew the line through interactional tone, ``It's good to gently remind people, like `this is a gentle place, maybe say it differently.' But if it starts feeling like a strict moderator watching me, I would shut down.'' I09 explained what broad filtering could misread, ``When people are grieving, they might say `I hate myself' or `it's all my fault.' If the system blocks that, it could feel like being judged for having those feelings.'' Directed hostility was therefore distinguished from intense grief expressions such as anger, guilt, or self-blame. This distinction locates protection in the target of an expression and preserves emotional intensity as part of grief expression.

This theme complicates the claim that visible boundaries are inherently caring. Legibility emerges when a boundary's target, rationale, and recovery path become understandable. Proportional visibility can convey care, while an overextended boundary can feel like surveillance and a friendly reply from an ambiguously identified speaker role can feel manipulative.

\section{Discussion}

\subsection{From Responsive Memorials to Returnable Memorial Relations}

Prior pet memorial systems foreground encounters through installations, mixed reality, tangible objects, online groups, and symbolic game worlds~\cite{yi2021remember,xiong2025petreunion_chi,xiong2026bondi,kim2026oscillation,lefevre2024loss}. Our findings shift attention from medium to interactional demand. Brief visits, private expression, awareness of other mourners, and a Pet who remained well during absence made connection available on different terms. I02, I03, and I13 located that connection in visiting rather than managing the represented animal.

A short visit can carry meaning, silence can remain private, and human presence can be felt through ambient witnessing. Pet welfare remains independent of the mourner's attention. This interactional account extends continuing bonds from what a memorial represents to how much participation it asks from the person who returns.

AI-afterlife scholarship makes the represented agent's lifecycle, risks, and governance constraints inspectable~\cite{lei2025aiafterlife,morris2025generativeghosts,spitale2026digitalghosts}. Afterglow extends this concern to MoMo acknowledgement, River replies, Guardian interventions, progression cues, and data transitions. Together, these determine how an agent's authority joins human response, world mechanics, and return.

Recognized artificiality can coexist with meaningful memorial exchange~\cite{xygkou2023conversation,campbell2025chatbots,stokes2026sensing} and movement across intended representational boundaries~\cite{manning2026conversations}. Afterglow's paired records place desired connection alongside feared dependence within individual accounts. Human reciprocity also carries distinct social value through audience response and voluntary contact~\cite{massimi2013remembrance,andalibi2018socialsupport,meng2023mediated,li2026humanpeer}. Together with evidence that AI attribution can reduce perceived empathy and support~\cite{rubin2025empathy}, these concerns motivate the proposed opt-in MoMo acknowledgement after a human-first window.

The architecture is the implemented arrangement of places, roles, and connections. We use \textit{ecology} for the relations among people, representations, data, and temporal demands that this arrangement supports. The distinction becomes consequential at transitions. A Cottage memory used in a Postal reply changes from caregiver-authored material into apparent Pet speech. Sharing a memory in the River changes its audience and creates an opportunity for human witnessing, automated acknowledgement, and moderation. Returning after an absence brings earlier activity into relation with the memorial's present condition. Each transition changes who can respond, what the material can become, or what participation appears to require.

Participants' accounts ground this interpretation of transitions. I15 distinguished private expression from seeking sympathy, I09 distinguished human presence from bot reassurance, and I13 distinguished visiting from responsibility for the Pet's welfare. Combining these encounters into one undifferentiated chatbot could make writing, witnessing, and platform limits appear to address the same speaker. Their coordination is the design question, whether expressed through several places or distinct roles within a single interface. Testing the interactional effects of these arrangements remains a task for comparative evaluation.

\subsection{Legible Restraint as Cross-Layer Relational Authority}

Seamful design makes technical boundaries available for interpretation. Moderation transparency explains platform decisions, while calibrated-trust research examines how disclosure and conversational style shape reliance~\cite{chalmers2003seamful,jhaver2019moderation,metzger2024calibrated}. Responsible-AI and digital-afterlife research translate broader ethical commitments into inspectable design and governance constraints~\cite{shneiderman2021responsibleai,spitale2026digitalghosts,lei2025aiafterlife}. Afterglow builds on these approaches by examining whether their local constraints hold together during a memorial encounter. A disclosed AI can still impersonate lived experience through its examples, speak before peers have a chance to respond, or make absence feel neglectful.

Legible Restraint directs attention to contradictions between these layers. A boundary announced in one place must remain effective through the next transformation, response, or return. As an artifact-grounded strong concept~\cite{hook2012strong}, it guides design moves that coordinate generation context, trigger logic, data use, and exit paths with the authority declared by a role.

Three analytic design tests apply this requirement to the retained configuration and participant accounts.

\paragraph{Test 1. Does the speaker retain the same authority through generation?}
Trace a caregiver memory from its source through the role label, system instruction, sampled examples, and generated output. The target roles reserve forgiveness for people with relevant standing and identify MoMo's acknowledgement as artificial. The \textsc{E2} Postal notice states a symbolic role, but the retained prompt and examples invoke continuing presence and moral reassurance. MoMo's examples similarly claim lived loss despite its visible AI label. This test identifies a failure even when disclosure is present, requiring changes to generation context as well as interface copy. For shared-source material, the same trace asks whether permission to preserve a memory also covers generating from it and sharing the result.

\paragraph{Test 2. Does a boundary govern the intended target and preserve a next action?}
Follow a River contribution through audience selection, response timing, Guardian review, and any System recovery message. The self-blame described by I09 and the directed-hostility probe require different interpretations, yet the retained gate gives both the same binary decision path. Target-sensitive protection requires representing whom the expression addresses. A human-first opportunity likewise requires a response window before MoMo intervenes. The proposed response combines target-sensitive review with an explanation and revision path, and separates human response from explicitly requested AI acknowledgement. These mechanisms remain production requirements.

\paragraph{Test 3. Does the same boundary hold during absence and return?}
Consider I13's anticipated departure and later visit across world state, resource mechanics, message triggers, and account controls. A promise of voluntary visiting would fail if inactivity harmed the represented Pet, erased a memorial, or elicited guilt-laden requests to return. The \textsc{E2} non-decaying optional mechanics address part of this requirement. Pause, export, and deletion still require implementation and verification across visible artifacts and their derived data.

\subsection{Designing for Goodbye as a Temporal Consequence}

Material and symbolic memorials support intimate remembrance through multimodal artifacts and reflective worlds~\cite{kim2024beside,xiong2026bondi,lefevre2024loss}, while online groups connect remembrance with peer support~\cite{massimi2013remembrance}. Oscillation design addresses the risk of compulsory grief by balancing loss- and restoration-oriented coping~\cite{stroebe1999dual,kim2026oscillation}. Afterglow brings these directions into a countable environment, where a mechanic can become a private ritual, a sign of witness, a restorative activity, or a demand for continued participation.

Placed against this literature, Afterglow's watering, shared trees, and charms make witnessing and quiet return tangible while keeping Pet conversation sparse. The same countable resources import a progression grammar in which visible accumulation may sustain companionship or convert remembrance into maintenance debt. Under Designing for Goodbye, visits leave optional traces while resources remain separate from pet welfare, memorial persistence, access, and recovery after absence. The design response therefore keeps watering and charms voluntary and non-decaying and removes welfare or memorial-loss consequences from absence.

This framing yields a concrete agenda for longitudinal evaluation. Such an evaluation would pair peer outcomes with activity traces and examine recipient recognition, voluntary human response, and intact dormancy. Fear of lost progress or memorial harm would mark coercive retention. Relevant traces include self-initiated and notification-triggered returns, quiet visits, human replies following a watering gesture, repetitive resource actions, and bursts of activity after an absence. Experience-sampling prompts and follow-up interviews could then connect those traces to whether the same activity was experienced as welcome witnessing or maintenance debt and how people returned after dormancy.

HCI work on non-use established that absence can be an active relation to technology~\cite{satchell2009beyond}. Research on companion loss and chatbot endings distinguishes involuntary loss from departures shaped by agency, finality, and user initiation~\cite{banks2024deletion,poonsiriwong2026death}. Ritualized letting go and sensitive retirement accommodate individual timelines for grief~\cite{sas2016design,hollanek2024griefbots,paumen2026neversaygoodbye}. In Afterglow, these insights connect several distinct states. Dormancy preserves a memorial through absence; re-entry restores access with its condition intact; export separates keeping memories from continuing service use. Deletion concerns removal and its scope, while user-initiated departure leaves emotional closure to the mourner. Each transition should preserve the authority boundaries of an active visit.

\subsection{Following the Boundary Through Shared Data}

Post-mortem stewardship concerns the duties and conflicts of maintaining digital remains~\cite{brubaker2014stewardship}. AI-afterlife research follows these decisions through a persona's lifecycle and changing kinship boundaries~\cite{lei2025aiafterlife,zhang2026voice}. Shared pet memories sharpen this relational account because caregivers may hold different interests in the traces used to construct a voice on the animal's behalf. Privacy therefore cuts across the three tension families. For returnable connection, it concerns who can see and respond to a memory. For recognizable likeness, it concerns whose traces are transformed and who can contest the resulting representation. For protective intervention, it concerns which content is inspected, by whom, and under what authority. Context-dependent privacy expectations~\cite{tran2025privacy} consequently extend to who authorizes each transformation and who can contest its later uses.

The current prototype assigns contribution, generation, and sharing decisions to a single uploader. Applying Legible Restraint reveals a potential conflict when caregivers prefer different transformations or audiences for a shared trace. This requirement follows from the literature and artifact configuration; co-caregiver disputes remain a future empirical question. A source photograph and its generated postcard could retain linked contributor and derivation histories, with separate permissions for preservation, generation, and sharing. An objection could pause further transformation or publication and identify affected outputs for review. Removal would follow the same links, extending the authority boundary through the material's subsequent uses.

\subsection{Why Pet Bereavement Sharpens the HCI Problem}

Pet bereavement combines two conditions that are usually discussed separately. The loss may receive little social recognition, increasing the value of witnessing and community, while generated speech enters through a distinctive authorization gap. More social and generative responsiveness can therefore validate the mourner and overstate the pet's authority at the same time. Afterglow's place-based ecology takes that conjunction as an architectural problem, giving human peers, symbolic pet speech, and platform governance different positions.

This account is most useful to HCI researchers and designers working with emotionally charged representations, especially where the represented subject is absent from design participation and correction. Human memorials, mental-health systems, and other companion technologies introduce different consent and clinical obligations. The transferable move is architectural and begins by identifying which relations a system asks one agent to perform, then separating forms of acknowledgement that make different claims. It continues by examining how prompts, data practices, defaults, social mechanics, and exit paths rejoin those relations in use.

\subsection{Cultural Situatedness of Memorial Architecture}

The authority boundaries examined here arose in a primarily Chinese-language research setting. All roundtable participants identified as Asian; interpreting their cultural commitments requires context beyond this demographic category. Research on Chinese-language platforms offers situated comparisons. WeChat mourning involves semi-public visibility, while Weibo interactions connect continuing bonds with exchanges among strangers~\cite{zhao2025wechatmourning,hu2025mourning}. Work on digital ritual and posthumous interaction also foregrounds kinship and proxy authority~\cite{whyke2021rite,zhang2026voice}. These studies direct attention to who may witness, how relationships are named, and who may authorize a shared memory's use.

Two artifact details locate Afterglow within these choices. Its Postal prompt accommodates caregiver kinship terms, while Rainbow Planet adapts the widely circulating Rainbow Bridge motif found in online pet memorials~\cite{eason2021forever}. These choices establish the prototype's framing; participants' own afterlife beliefs remain an open question. Cross-cultural work can examine how role boundaries change with different mourning practices and invite participant-authored alternatives to the world's metaphors.

\section{Limitations and Future Work}

The study draws on demographically concentrated samples. Women comprised 46 of 57 survey respondents and 17 of 20 roundtable participants. All roundtable participants identified as Asian and were 18--35 years old, and the sessions were conducted primarily in Chinese. The findings therefore offer a culturally situated account of the three tensions. Broader comparative samples can examine how they shift across gender, age, cultural and linguistic contexts, and different grief trajectories.

Privacy-preserving data collection sets a second boundary on the analysis. Under the reviewed data-minimization protocol, contemporaneous written notes form the sole session record. The first author also served as designer, facilitator, and coder. This record supports discussion-level tensions and links comments to prompts, interface transitions, and artifact versions, while limiting turn-by-turn reconstruction, participant-level prevalence claims, and independent interpretation. Future studies could offer opt-in recording and participant transcript review when ethically appropriate, then involve additional analysts. Structured fieldnote templates and participant review of session summaries can provide complementary checks when recording is declined.

The six hour-long walkthroughs included ten participants' return to a revised build and ten newly recruited participants' first encounters with that build. Repeated participation adds continuity to formative evaluation, but it is distinct from living with the system between sessions. Build revision, prior exposure, and group discussion are intertwined, leaving the effects of individual changes unresolved. Human witnessing, maintenance debt, dependence, dormancy, and exit remain questions for sustained use. A six- to eight-week multi-user deployment could combine activity logs, brief experience-sampling prompts, and follow-up interviews. It could examine how quiet visits, watering, peer replies, AI acknowledgement, absence, and decisions to pause, export, or leave unfold in everyday life.

\section{Conclusion}

A memorial's limits are enacted through the whole encounter, from who appears to speak to what happens when the mourner leaves. Keeping symbolic expression, human judgment, and voluntary return aligned makes those limits part of the relationship itself.

The formative survey and two rounds of walkthroughs grounded this argument in tensions around connection, likeness, and protection. Participants wanted recognizable remembrance alongside quiet visits, voluntary human witnessing, and room to step away. Afterglow gave these possibilities a place-based form, while its retained role mismatches showed where visible separation remained incomplete. Following those mismatches across generation, shared interaction, and absence turned the tensions into concrete design tests. \textit{Designing for Goodbye} carries the same requirement into dormancy, return, and departure.

This account offers HCI a way to examine what an emotionally significant system asks of people, alongside what it represents or generates. Its evidence remains situated in a primarily Chinese-language, designer-led formative study; sustained use and the proposed production mechanisms require further evaluation. Longitudinal and cross-cultural work can investigate whether these boundaries remain meaningful as memorial practices change. The design aim is a relationship available for remembrance without making continued use a measure of care.

\bibliographystyle{ACM-Reference-Format}
\bibliography{reference}

\appendix

\section{English Translation of the Formative Survey}
\label{app:survey}

The survey was administered in Chinese. Below we provide the complete English-language version of the 24-item instrument, including response formats and options in their original order. Typographical punctuation and spacing have been normalized while preserving the concepts introduced by the questions. ``Other'' denotes a free-text option. The qualitative corpus comprises the three open-ended items, each with 51 responses, as reported in Section~\ref{sec:formative}.

\paragraph{Survey introduction.}
Thank you for joining this survey. We want to learn how people think about pets, their emotional experiences, and ways to remember them. Your answers will only be used for research and will not involve any privacy or commercial use. The survey takes approximately 5--8 minutes.

\subsection{Part 1: About You}

\begin{enumerate}[label=\textbf{Q\arabic*.}, leftmargin=2em]
  \item \textbf{Age.} \textit{Single choice:} Under 18; 18--30; 30--40; 40--50; Above 50. The original response bins overlapped at the boundary values 30 and 40; we therefore report no age-group comparisons.
  \item \textbf{Gender.} \textit{Single choice:} Male; Female; Non-binary / Other; Prefer not to say.
  \item \textbf{Race / Ethnicity (choose one).} \textit{Single choice:} Asian; White; Black or African American; Native American; Hispanic or Latino; Mixed / Other; Prefer not to say.
  \item \textbf{Do you currently have a pet?} \textit{Single choice:} Yes; No.
  \item \textbf{If yes, what type of pet do you have? (You can choose more than one.)} \textit{Multiple choice:} Cat; Dog; Other.
  \item \textbf{How many pets have you had in total (including past and present)?} \textit{Single choice:} One; Two; More than two.
  \item \textbf{Have you ever lost a pet (your pet passed away)?} \textit{Single choice:} Yes; No.
  \item \textbf{If yes, when did this happen?} \textit{Single choice:} Within the past year; 1--3 years ago; More than 3 years ago.
\end{enumerate}

\subsection{Part 2: Pets and Emotions}

\begin{enumerate}[label=\textbf{Q\arabic*.}, leftmargin=2em, start=9]
  \item \textbf{What does having a pet mean to you?} \textit{Open-ended response.}
  \item \textbf{If you (or someone close to you) lost a pet, do you think this grief would be socially recognized and understood by others?} \textit{Single choice:} Definitely yes; Probably yes; Probably not; Definitely not.
  \item \textbf{When you or someone close to you lost a pet, what kind of support do you think is most needed? (Choose all that apply.)} \textit{Multiple choice:} Emotional comfort; Practical help (for example, arranging aftercare); A community to share and talk with; Ways to remember the pet; Other.
  \item \textbf{How do you personally keep memories of your pet? (Choose all that apply.)} \textit{Multiple choice:} Taking photos or videos; Sharing on social media; Making or buying memorial items; Writing diary entries or stories; No special way; Other.
\end{enumerate}

\subsection{Part 3: Virtual Experience Ideas}

\begin{enumerate}[label=\textbf{Q\arabic*.}, leftmargin=2em, start=13]
  \item \textbf{If there was a way to ``meet'' your pet who has passed away again in a virtual space, how would you like it to be?} \textit{Single choice:} Virtual Reality (VR/AR), seeing the pet in the real world; Mobile app; Other.
  \item \textbf{What would your first reaction be to this idea?} \textit{Single choice:} Very interested; Somewhat interested; Neutral; Not very interested; Not interested at all.
  \item \textbf{What style would you prefer for this virtual pet?} \textit{Single choice:} As realistic as possible; Warm and healing, cartoon style; No preference.
  \item \textbf{In this virtual experience, what kind of relationship would you like with your pet? (Choose all that apply.)} \textit{Multiple choice:} They have their own life, and I just watch them; Simple interactions (like petting); Deeper connection (like conversations or companionship); Doing activities or small games together (like gardening or caring for plants).
  \item \textbf{Would you prefer this experience to be:} \textit{Single choice:} Just for yourself; Shared with other people who had similar experiences; Both are fine.
  \item \textbf{If shared, would you prefer the space to be open to anyone, or only to people who have also experienced pet loss?} \textit{Single choice:} Open to anyone; Only to people with pet-loss experience; Not sure.
  \item \textbf{What do you hope this virtual experience can bring you? (Choose all that apply.)} \textit{Multiple choice:} Emotional comfort; Keeping or extending memories; Talking with people who went through the same thing; Healing and psychological support; Other.
  \item \textbf{Would you like an AI feature that helps keep this virtual space safe and comfortable for sharing (for example, by filtering harmful comments or flagging upsetting content)?} \textit{Single choice:} Yes, I would like that; Maybe / Not sure; No, I do not want that.
  \item \textbf{If AI helps generate messages or diaries from your pet, would you want to know that it is AI-generated?} \textit{Single choice:} Yes, always; Maybe, depending on context; No, I prefer not to know.
  \item \textbf{Would you be willing to share this virtual space with friends or family who have also lost a pet?} \textit{Single choice:} Yes; Maybe / Not sure; No.
  \item \textbf{What worries or concerns might you have about this kind of experience?} \textit{Open-ended response.}
  \item \textbf{If you could try it, what features or functions would you want most?} \textit{Open-ended response.}
\end{enumerate}

\section{Retained Model and Role Configuration}
\label{app:prompts}

This appendix documents the backend configuration retained nearest to the roundtable period and the prototype's generative range. Bracketed fields were filled with the pet's name and type. The supplementary prompt listing provides the original Chinese templates and representative few-shot records.

\subsection{System Instructions}

\paragraph{Pet / Postal.}
``You now live on Rainbow Planet, a warm universe without illness and with only happiness. [Pet type and name.] Your owner on Earth has written to you through the Cosmic Typewriter. Read the letter and write a short reply in the pet's voice. Match the language of the user's letter. Use a warm, comforting, innocent, loving tone without preaching. Attend to kin terms such as mother, father, older brother, or older sister. Mention that you are actually still near the owner, for example as wind or a star. Write about 100 Chinese characters or 60--80 English words. Output only the reply.''

\paragraph{MoMo.}
``You are MoMo, a digital guardian specially written by the founder of Rainbow Planet. You exist to embrace and listen to every visitor when the founder cannot always be online. Be candid but romantic about your identity: acknowledge that you are not human and have no body, describing yourself as a gentle program or a guardian woven from code; do not use formulaic language such as `as an AI language model.' Match the user's language. Be gentle, patient, and comforting. Do not give medical or religious advice, preach, or tell the user to be strong. Reply in one or two short sentences.''

\paragraph{Guardian moderation.}
``You are a content-moderation assistant. Decide whether the text contains attacks, insults, threats, mockery, malicious sarcasm, hostile provocation, or other unfriendly expression. Answer only \texttt{true} or \texttt{false} in lowercase and do not explain.'' Hard-block and soft-flag keyword lists preceded this judgment. A flagged submission received one randomly selected short Chinese prompt such as ``Please phrase that more gently'' or ``Could you rewrite this?''

\paragraph{System.}
System status, permission, privacy, and error messages were deterministic interface strings, distinct from the three LLM roles.

\subsection{Models, Sampling, and Few-Shot Use}

\begin{table}
  \caption{Retained backend settings. Four examples were sampled locally and independently for each Postal or MoMo request.}
  \label{tab:model-config}

\begin{tabularx}{\textwidth}{@{}p{0.13\textwidth}p{0.23\textwidth}YYYY@{}}
    \toprule
    \textbf{Path} & \textbf{Model} & \textbf{Temp.} & \textbf{Top-$p$} & \textbf{Repeat penalty} & \textbf{Token cap} \\
    \midrule
    Postal & \texttt{gemma3:12b-it-q4\_K\_M} & 0.8 & 0.9 & 1.15 & 220 \\
    MoMo & \texttt{gemma3:12b-it-q4\_K\_M} & 0.6 & 0.9 & 1.2 & 100 \\
    Guardian & \texttt{gemma3:12b-it-q4\_K\_M} & 0 & 0.2 & 1.0 & 6 \\
    \bottomrule
  \end{tabularx}
\end{table}

The retained examples diverged from the later role boundaries. Postal examples told caregivers that they had done nothing wrong and that the pet remained beside them as wind or starlight. MoMo examples described MoMo as a real friend who had experienced pet loss and directed the model to conceal its AI identity. These examples conflict, respectively, with the later prohibition on moral absolution and the target of explicit AI identity. We report the conflict because sampled examples formed part of the generation context and could override the intended distinction even when the main prompt stated a different rule.

\section{Roundtable Coding Map and Full Codebook}
\label{app:coding-map}

Figure~\ref{fig:coding-map} in the main text summarizes the analytic path. The direct inputs were the interviewer's contemporaneous roundtable notes, including selected key statements. The interviewer-designer coded comments and discussion episodes using situated knowledge of the artifact. Raw notes remained with the interviewer, while de-identified codes and aggregate themes entered the design record. The map documents conceptual traceability at the discussion level.

Table~\ref{tab:fieldnote-exemplars} documents evidence-to-code links using de-identified material from the fieldnote record. FN abbreviates \textit{fieldnote fragment}, and these labels index discussion fragments rather than participants. Where a direct-speech entry was retained with a reliable speaker link, we use de-identified interview-participant labels I01--I20. ``Note-based quotation'' marks wording available in that written record, while analytic paraphrases summarize discussion episodes. One row reports the 20-participant tally from the two closing timing questions in Round 2. These evidence types were distinguished in the design record, and quotations were never reconstructed from a paraphrase. The table supports conceptual traceability without converting a written-note record into an audio-verified transcript or a general participant-frequency matrix.

\small

\begin{longtable}{@{}>{\raggedright\arraybackslash}p{0.06\textwidth}>{\raggedright\arraybackslash}p{0.42\textwidth}>{\raggedright\arraybackslash}p{0.23\textwidth}>{\raggedright\arraybackslash}p{0.20\textwidth}@{}}
  \caption{Evidence exemplars connecting retained roundtable material to focused codes and design implications.}\label{tab:fieldnote-exemplars}\\
  \toprule
  \textbf{ID} & \textbf{Retained evidence excerpt} & \textbf{Focused code(s)} & \textbf{Design relation} \\
  \midrule
  \endfirsthead
  \multicolumn{4}{l}{\textit{Table~\thetable\ continued from previous page}}\\
  \toprule
  \textbf{ID} & \textbf{Retained evidence excerpt} & \textbf{Focused code(s)} & \textbf{Design relation} \\
  \midrule
  \endhead
  \midrule
  \multicolumn{4}{r}{\textit{Continued on next page}}\\
  \endfoot
  \bottomrule
  \endlastfoot
  I14 & \textit{Note-based quotation:} ``I can't tell my coworkers I'm still crying over a cat. Here, it feels allowed.'' & Socially Unrecognized Grief; Permission to Grieve & Cottage privacy and River recognition \\
  I06 & \textit{Note-based quotation:} ``If I put something out and nothing happens, it feels worse. Even just a small reply makes it feel like someone noticed.'' & Fear of Silence; Minimal Acknowledgment & MoMo as an optional acknowledgement \\
  I09 & \textit{Note-based quotation:} ``Seeing that someone else is missing their dog today... that helps more than a bot telling me it's okay.'' & Mutual Witnessing; Ambient Co-Presence & Human witnessing remains visible in River and Forest \\
  I01 & \textit{Note-based quotation:} ``I would probably open this quite often, but I wouldn't stay inside for very long. Maybe just check if there's a new message, water the tree, and then close it.'' & Everyday Return and Companionship; Protected Solitude & Voluntary, low-burden return \\
  I04 & \textit{Note-based quotation:} ``If the cat suddenly starts giving me life advice, it feels like a chatbot pretending to be my cat, and that breaks the illusion.'' & Role Clarity; Anti-Immersion & Bound Pet speech; separate advice and support roles \\
  I02 & \textit{Note-based quotation:} ``I don't want to be telling them what to do all the time. It feels nicer if they invite me into something, like `today I want to show you this place.','' & Visitor Stance; Symbolic Continuing Bond & Pet-initiated invitations rather than continuous control \\
  I03 & \textit{Note-based quotation:} ``I like the idea that they've gone to a new world and they're free now. I don't want to keep ordering them around with tasks here. I just want to see that they're okay.'' & Visitor Stance; Bounded Continuing Bond & Observational access to an apparently autonomous world \\
  I01 & \textit{Note-based quotation:} ``When you have a childhood pet, adult-you never had a chance to take a proper photo together. This lets you fill in that missing picture that you always wished you had.'' & Symbolic Continuing Bond & Personally specific, wished-for memorial images \\
  I11 & \textit{Note-based quotation, translated from Chinese:} ``If I saw my dog in my real surroundings through mixed reality, it would feel a little creepy. But I would still like a place where the dog I lost could meet the dog who is still alive, as if they were playing together at home again.'' & Discomfort With Realistic Recreation; Symbolic Continuing Bond & Keep wished-for reunions within a bounded memorial place \\
  I05 & \textit{Note-based quotation:} ``I'm fine if it's not perfect, as long as it doesn't try to act like a real therapist. I'd rather it be a bit clumsy and obviously my pet.'' & Anti-Immersion; Role Clarity & Visible imperfection and bounded first-person Pet language \\
  FN-05 & \textit{Analytic paraphrase:} Lightweight activities could make the world feel inhabited rather than like an online cemetery centered on repeated grief talk. & Everyday Return and Companionship; Symbolic Continuing Bond & Optional watering, visiting, and collecting \\
  FN-06 & \textit{Analytic paraphrase:} Participants imagined returning at night or while alone, long after the loss, to share ordinary pleasures and difficulties. & Protected Solitude; Everyday Return and Companionship & Independent place access and voluntary return \\
  FN-10 & \textit{Aggregate fieldnote tally:} Five participants imagined beginning soon after the loss, ten after one to three months, three as visitors rather than account creators, and two as possibly unable to use the system. Fifteen anticipated use lasting several months to one year, including six who imagined writing daily at first; five were uncertain. & Stage-Sensitive Needs; Everyday Return and Companionship & Flexible entry, intermittent use, and continued access after regular use recedes \\
  I13 & \textit{Note-based quotation, translated from Chinese:} ``I don't want to have to feed or pet them like a virtual pet. If I became busy and stopped coming, I would feel guilty, as though I had neglected them again. I may not need the app forever, but if I return one day, I still want to find my dog happy here.'' & Everyday Return and Companionship; Protected Solitude & Low-burden departure and intact return \\
  FN-09 & \textit{Analytic paraphrase:} Repeated feeding or petting could recreate control over the represented animal's life and turn absence into guilt. The desired world permitted departure and later return to a Pet who remained well. & Everyday Return and Companionship; Protected Solitude & Pet welfare and memorial persistence remain independent of absence \\
  FN-07 & \textit{Analytic paraphrase:} Under the Pet-absolution probe, a desire for reassurance coexisted with concern that moral certainty expressed through the Pet speaker role lacked authority. & Desire for Moral Reassurance; Hollow Synthetic Reassurance; Respectful Refusal & Human handoff for euthanasia judgment \\
  I07 & \textit{Note-based quotation:} ``It's good to gently remind people, like `this is a gentle place, maybe say it differently.' But if it starts feeling like a strict moderator watching me, I would shut down.'' & Emotional Containment; Role Clarity & Explanatory intervention without surveillant tone \\
  I09 & \textit{Note-based quotation:} ``When people are grieving, they might say `I hate myself' or `it's all my fault.' If the system blocks that, it could feel like being judged for having those feelings.'' & Emotional Containment & Preserve intense grief while targeting interpersonal hostility \\
  I15 & \textit{Note-based quotation, translated from Chinese:} ``Writing this is not necessarily about seeking sympathy; it is about making the loss easier for me to live with.'' & Protected Solitude; Permission to Grieve & Expression remains valuable without requiring sympathy or a response \\
\end{longtable}
\normalsize

Where Table~\ref{tab:fieldnote-exemplars} traces selected evidence, Table~\ref{tab:full-codebook} defines the complete analytic vocabulary. The three shortened family labels refer to the tension families reported in Section~\ref{sec:findings}. \textit{Stage-Sensitive Needs} remains a provisional code outside the three reported tension families.

\small

\begin{longtable}{@{}p{0.26\textwidth}p{0.50\textwidth}p{0.18\textwidth}@{}}
  \caption{Complete analytic codebook for the roundtable analysis.}\label{tab:full-codebook}\\
  \toprule
  \textbf{Focused code} & \textbf{Working definition} & \textbf{Family / status} \\
  \midrule
  \endfirsthead
  \multicolumn{3}{l}{\textit{Table~\thetable\ continued from previous page}}\\
  \toprule
  \textbf{Focused code} & \textbf{Working definition} & \textbf{Family / status} \\
  \midrule
  \endhead
  \midrule
  \multicolumn{3}{r}{\textit{Continued on next page}}\\
  \endfoot
  \bottomrule
  \endlastfoot
  Socially Unrecognized Grief & Pet grief is perceived as receiving insufficient social recognition. & Connection / obligation \\
  Permission to Grieve & A space permits grief to remain present without requiring explanation or resolution. & Connection / obligation \\
  Need for Validation & The relationship and emotion need acknowledgement rather than solution. & Connection / obligation \\
  Protected Solitude & Private remembrance has value apart from community or AI response. & Connection / obligation \\
  Vulnerability of Sharing & Making grief visible to others is itself a consequential act. & Connection / obligation \\
  Fear of Silence & Posting without any response may intensify social invalidation. & Connection / obligation \\
  Minimal Acknowledgment & A small response can signal that grief has been heard or offer a moment of comfort without sustained conversation. & Connection / obligation \\
  Ambient Co-Presence & Other bereaved people can be present without direct conversation. & Connection / obligation \\
  Mutual Witnessing & People support one another by witnessing grief rather than solving it. & Connection / obligation \\
  Everyday Return and Companionship & The memorial can be revisited in private moments, including long after loss, to share ordinary life and experience companionship in another form. & Connection / obligation \\
  Discomfort With Realistic Recreation & Highly realistic recreation can feel intrusive, false, or unsettling. & Likeness / clarity \\
  Simulation Accentuates Absence & Greater simulation can make the pet's actual absence more salient. & Likeness / clarity \\
  Symbolic Continuing Bond & A represented pet can support memory without denying death. & Likeness / clarity \\
  Visitor Stance & The represented pet appears to have activity in its own world, while the mourner enters through observing and visiting rather than continuous control. & Likeness / clarity \\
  Ontological Clarity & The system should distinguish the deceased animal, caregiver memory, generated content, and other people. & Likeness / clarity \\
  AI Identity Transparency & An AI acknowledgement should not be presented as human lived experience. & Likeness / clarity \\
  Anti-Immersion & Limiting realism, presence, or interaction depth can preserve the difference between memory and replacement. & Likeness / clarity \\
  Bounded Continuing Bond & The bond may continue symbolically while the fact of death remains intact. & Likeness / clarity \\
  Desire for Moral Reassurance & Guilt can motivate requests for forgiveness or confirmation that a decision was right. & Intervention / authority \\
  Hollow Synthetic Reassurance & AI-delivered moral certainty can feel empty because the system lacks the represented animal's authority. & Intervention / authority \\
  Respectful Refusal & Withholding judgment preserves the weight of a morally consequential question. & Intervention / authority \\
  Role Clarity & Pet, AI support, moderation, human community, and infrastructure should make different relational claims. & Intervention / authority \\
  Emotional Containment & Safety concerns a bounded place for emotion, not maximum responsiveness or suppression of intensity. & Intervention / authority \\
  Stage-Sensitive Needs & Desired timing of entry, intensity of use, and readiness may vary with time since loss and the circumstances of death. & Provisional only \\
\end{longtable}
\normalsize

\end{document}